\documentclass[aps,prd,amsmath,showpacs,preprintnumbers,
               twocolumn]{revtex4}
\usepackage{amsmath}
\usepackage{graphicx}
\usepackage{latexsym}

\newcommand{\beqa}{\begin{eqnarray}}
\newcommand{\eeqa}{\end{eqnarray}}
\newcommand{\beq}{\begin{equation}}
\newcommand{\eeq}{\end{equation}}

\newcommand{\<}{\langle}
\renewcommand{\>}{\rangle}
\newcommand{\tr}{\mathop{\rm Tr}\nolimits}
\newcommand{\1}{1\!\!\!\bot}
\def\reff#1{(\ref{#1})}

\def\F2{\langle \alpha_s F_{\mu\nu} F^{\mu\nu} \rangle}

\def\A2{\langle A_{\mu}^a A_{\mu}^a \rangle}

\def\gc{\left  \langle f^{bcd} \overline{\eta}^{c} \eta^{d}            \right \rangle}

\def\gca{\left \langle f^{bcd} \eta^{c}            \eta^{d}            \right \rangle}

\def\gcb{\left \langle f^{bcd} \overline{\eta}^{c} \overline{\eta}^{d} \right \rangle}

\def\Z1{\widetilde{Z}_1}

\def\spose#1{\hbox to 0pt{#1\hss}}
\def\ltapprox{\mathrel{\spose{\lower 3pt\hbox{$\mathchar"218$}}
 \raise 2.0pt\hbox{$\mathchar"13C$}}}
\def\gtapprox{\mathrel{\spose{\lower 3pt\hbox{$\mathchar"218$}}
 \raise 2.0pt\hbox{$\mathchar"13E$}}}

\begin{document}

\title{Ghost condensation on the lattice}
\author{Attilio Cucchieri, Tereza Mendes and Antonio Mihara}
\affiliation{Instituto de F\'\i sica de S\~ao Carlos, Universidade de S\~ao Paulo, \\
C.P.\ 369, 13560-970, S\~ao Carlos, SP, Brazil}


\begin{abstract} 
\noindent
We perform a numerical study of ghost condensation --- in the so-called
Overhauser channel --- for $SU(2)$ lattice gauge theory in minimal Landau
gauge. The off-diagonal components of the momentum-space ghost propagator
$G^{cd}(p)$ are evaluated for lattice volumes $V = 8^4$, $12^4$, $16^4$, $20^4$,
$24^4$ and for three values of the lattice coupling: $\beta = 2.2$, $2.3$, $2.4$.
Our data show that the quantity $\phi^b(p) = \epsilon^{bcd}
G^{cd}(p) / 2$ is zero within error bars, being characterized by very
large statistical fluctuations. On the contrary, $|\phi^b(p)|$
has relatively small error bars and behaves at small momenta as 
$\, L^{-2} p^{-z} \,$,
where $L$ is the lattice side in physical units and $z \approx 4$. 
We argue that the large fluctuations for $\phi^b(p)$ come from
spontaneous breaking of a global symmetry and are 
associated with ghost condensation. It may thus be necessary 
(in numerical simulations at finite volume) to consider 
$|\phi^b(p)|$ instead of $\phi^b(p)$, 
to avoid a null average due to tunneling between different broken vacua. 
Also, we show that $\phi^b(p)$ is proportional to the Fourier-transformed 
gluon field components ${\widetilde A}_{\mu}^b(q)$. This
explains the $L^{-2}$ dependence of $|\phi^b(p)|$,
as induced by the behavior of $| {\widetilde A}_{\mu}^b(q) |$.
We fit our data for $|\phi^b(p)|$
to the theoretical prediction $( r / L^2 + v ) / (p^4 + v^2)$, obtaining for the
ghost condensate $v$ an upper bound of about $0.058$ GeV$^2$.
In order to check if $v$ is nonzero in the continuum limit, one probably needs
numerical simulations at much larger physical volumes than the ones we consider.
As a by-product of our analysis,
we perform a careful study of the color structure of the inverse
Faddeev-Popov matrix in momentum space.

\end{abstract}

\pacs{11.15.Ha,   
      12.38.Aw,   
      12.38.Gc,   
      14.80.-j    
} 

\maketitle


\section{Introduction}

The QCD vacuum is known to be highly non-trivial at low energies \cite{Shuryak}.
This non-trivial structure manifests itself through the
appearance of vacuum condensates, i.e.\ vacuum expectation values of certain
local operators. In perturbation theory these condensates vanish,
but in the SVZ-sum-rule approach \cite{Shifman:1978bx,Shifman:1978by}
they are included as a parametrization of non-perturbative effects
in the evaluation of phenomenological quantities.
Of course, the local operators considered must be gauge invariant in order to be added
as higher-dimension operators (higher twists) to the usual perturbative expansion of physical
quantities. The two main such operators are $\,\alpha_s F_{\mu\nu} F^{\mu\nu} \,$ and
$\,m_q \overline{\psi}_q \psi_q$; their vacuum expectation values are the so-called
gluon and quark condensates. Both of these operators have mass dimension four.

\vskip 2mm

In recent years, (gauge-dependent) condensates of mass dimension two have also received
considerable attention \cite{Gubarev:2000eu,Gubarev:2000nz,Dudal:2002ye,
Lemes:2002ey,Lemes:2002jv,Lemes:2002rc,Dudal:2002xe,Dudal:2002pq,Dudal:2003gu,
Dudal:2003dp,Dudal:2003pe,Dudal:2003np,Dudal:2003by,Sobreiro:2004us,
Dudal:2004rx,Dudal:2004ch,Dudal:2005bk,Dudal:2005na,Dudal:2005zr,Capri:2005,
Kondo:2000ey,Kondo:2001kh,Kondo:2001nq,Kondo:2001tm,Kondo:2002cy,Kondo:2003uq,
Kondo:2003sy,Kondo:2005zs,Slavnov:2004rz,Slavnov:2005av,Bykov:2005tx}.
In particular, the gauge condensate $\langle A_{\mu}^b A_{\mu}^b \rangle$
has been largely studied, since it should be sensitive to topological 
structures such as monopoles \cite{Gubarev:2000eu,Gubarev:2000nz}. Thus,
this parameter could play an important role in the quark-confinement scenario
through monopole condensation \cite{'tHooft:1981ht,Suzuki:1992rw,Chernodub:1997dr}.
Moreover, the existence of a gauge condensate would imply a dynamical
mass generation for the gluon and ghost fields. This result has already
been obtained in various gauges \cite{Dudal:2003by,Sobreiro:2004us,Dudal:2005na,
Kondo:2002cy,Kondo:2003sy}, but clearly, since the
operator $ A_{\mu}^b A_{\mu}^b $ is not gauge invariant, it is difficult to assign
a physical interpretation to the condensate $\langle A_{\mu}^b A_{\mu}^b \rangle$.
Recently, evidence that the expectation value $\langle A_{\mu}^b A_{\mu}^b \rangle$
may be gauge independent was presented in \cite{Kondo:2005zs,Slavnov:2004rz,Slavnov:2005av,Bykov:2005tx}.
At the same time, the authors of Refs.\ \cite{Gubarev:2000eu,Gubarev:2000nz} have
shown that the minimal value of $\langle A_{\mu}^b A_{\mu}^b \rangle$ (on each gauge orbit) has a
gauge-invariant meaning. Let us note that this minimal value is achieved
by considering absolute minima in the minimal Landau gauge,
i.e.\ for configurations belonging to the so-called (Landau) fundamental modular
region \cite{Zwanziger:1993dh}.

\vskip 2mm

Other vacuum condensates of mass dimension two considered by several groups
are the ghost condensates $\gc$, $\gca$ and $\gcb$. These
condensates were first introduced in $SU(2)$ gauge theory in maximally
Abelian gauge (MAG) \cite{Schaden:1999ew,Schaden:2000fv,Schaden:2001xu,
Lemes:2002ey,Dudal:2002xe,Kondo:2000ey,Kondo:2003uq}.
More recently, the same condensates have been studied in other gauges
\cite{Dudal:2002ye,Lemes:2002jv,Lemes:2002rc,Dudal:2003dp,Kondo:2001tm,Kondo:2001kh}, such as 
the Curci-Ferrari and the Landau gauges.
In all cases it was found that the ghost condensates are related to
the breakdown of a global $SL(2,R)$ symmetry \cite{Dudal:2002ye,Alkofer:2000wg}.
In MAG the diagonal and off-diagonal components of the ghost propagators
are modified \cite{Lemes:2002ey,Kondo:2000ey} by ghost condensation.
Similar results have been obtained in other gauges \cite{Dudal:2003dp,Kondo:2001tm}.
In particular, in Landau gauge it was shown \cite{Dudal:2003dp}
that the off-diagonal (anti-symmetric) components of the ghost propagator $G^{cd}(p)$
 --- or, equivalently, the quantity $\phi^b(p)$ --- 
are proportional to the ghost condensate $v$. In the same reference
it is argued that the breaking of the global
$SU(N_c)$ color symmetry, related to this non-zero expectation value for $G^{cd}(p)$,
may occur only in the unphysical sector of the Hilbert space and
therefore should not be (explicitly) observable in the physical subspace.
Very recently \cite{Capri:2005}, the effects of the ghost condensate
$\gc$ and of the gauge condensate $\langle A_{\mu}^b A_{\mu}^b \rangle$
have been considered together for the $SU(2)$ case in Landau gauge, showing that
ghost condensation induces a splitting in the gluon mass related to the 
gauge condensate.

It is interesting that the existence of different ghost condensates
has an analogue in the theory of superconductivity \cite{Dudal:2003dp}.
In particular, the condensates $\gca$ and $\gcb$ correspond to the
so-called BCS effect \cite{Bardeen:1957kj,Bardeen:1957mv},
in which particle-particle and hole-hole pairing occur,
while the condensate $\gc$ is analogous to the Overhauser effect \cite{Over},
in which particle-hole pairing occurs. Also, in the Curci-Ferrari
gauge, both effects have to be considered \cite{Dudal:2003dp} in order to obtain
an action invariant under the so-called Nakanishi-Ojima (NO) algebra. It is this
NO algebra that includes an $SL(2,R)$ sub-algebra.

\vskip 2mm

Finally, a mixed gluon-ghost condensate of mass dimension two has also been studied
by several authors \cite{Kondo:2001nq,Kondo:2001tm,Dudal:2002xe,Dudal:2003gu,
Dudal:2004rx,Dudal:2005bk}, using various gauges. This mixed condensate is of
particular interest when considering interpolating gauges \cite{Dudal:2004rx}.
Indeed, it allows to generalize and relate results obtained in different gauges
for the gauge condensate $\langle A_{\mu}^b A_{\mu}^b \rangle$ and the ghost condensates.
Moreover, in MAG this mixed condensate would
induce a dynamic mass for the off-diagonal gluons \cite{Dudal:2005bk},
giving support to the Abelian-dominance scenario \cite{Ezawa:1982bf,
Suzuki:1988qa,Amemiya:1998jz,Bornyakov:2003ee}. Thus,
the various gauge and ghost condensates could all play an
important role in the dual superconducting scenario of
quark confinement \cite{Nambu:1974zg,Mandelstam:1974pi,Suganuma:2003ds},
being related to monopole condensation and to Abelian dominance.

\vskip 2mm

Possible effects of the gauge condensate $\langle A_{\mu}^b A_{\mu}^b \rangle$
on propagators and vertices (in Landau gauge) have been studied through
lattice simulations in Refs.\
\cite{Boucaud:2001st,Boucaud:2002nc,Boucaud:2002jt,RuizArriola:2004en,Furui:2005bu,
Boucaud:2005rm,Boucaud:2005ce}, yielding $\langle A_{\mu}^b A_{\mu}^b \rangle \approx 3$ GeV$^2$.
On the other hand, the ghost condensate has been
considered numerically only in Ref.\ \cite{Furui:2003jr}, for minimal Landau gauge
in the Overhauser channel, both for the $SU(2)$ and $SU(3)$ cases. The authors find that
the color off-diagonal anti-symmetric components of the ghost propagators are identically zero,
while the symmetric components are zero within error bars, but
have fluctuations proportional to $N^{-2} \hat{p}^{-4}$, where $\hat{p}$ is
the lattice momentum and $N$ is the lattice side.

\vskip 2mm

In this paper we carry out a thorough investigation of ghost condensation in the
Overhauser channel for pure $SU(2)$ Yang-Mills theory in minimal Landau gauge.
In particular, we evaluate numerically the off-diagonal components of the ghost
propagator $G^{cd}(p)$ as a function of the momentum $p$. We find that
$ \langle \phi^b(p) \rangle = \epsilon^{bcd}\, \langle G^{cd}(p) \rangle / 2 $
is zero within error bars, but with large fluctuations. On the other hand, if
we consider the absolute value of $\phi^b(p)$, we find that $\,\langle \, \left|
\, \phi^b(p) \, \right| \, \rangle \, $ behaves at small momenta as $\, L^{-2}
p^{-z} \,$, where $L$ and $p$ are, respectively, the lattice side and the momentum
in physical units and $z \approx 4$. As we will show [see Eq.\ \reff{eq:phimom}],
the sign of $\phi^b(p)$ is
related to the sign of the Fourier-transformed gluon field components
${\widetilde A}(q)$. 
At the same time, one should recall that a nonzero value for $\phi^b(p)$
is related to the spontaneous breaking of a global $SL(2,R)$ symmetry
\cite{Dudal:2002ye,Alkofer:2000wg}. Thus, the
situation here is similar to the one encountered in numerical studies of 
spin systems with
nonzero spontaneous magnetization, in the absence of an external magnetic field.
Indeed, since spontaneous symmetry breaking can occur only in the thermodynamic limit,
when performing numerical simulations at finite volume one always finds (after
sufficiently many Monte Carlo steps) a null magnetization for all nonzero
temperatures~\footnote{See for example \cite[Sections 2.3.3 and 2.3.4]{Binder-Heermann}
and \cite[Section 2.2.5]{Mouritsen}.}.
This is due to the fact that, at finite volume, the probability that the system
may pass from a state near a broken vacuum to a state near a different broken vacuum
is always nonzero.
To avoid this problem, the usual solution is to consider the absolute value of
the magnetization or the root mean square order parameter, i.e.\
$\langle \, | \, \vec{M} \, | \, \rangle$ or $\,\langle\,\vec{M}^2\,\rangle^{\!\,1/2}\,$.
The results obtained using these two quantities do not agree at finite volume, but
they produce the same infinite-volume-limit result \cite[Fig. 2.13]{Binder-Heermann}.
Alternatively, one can also simulate spin models with an external magnetic
field $H$ \cite{Engels:1999wf} and then consider the limit $H \to 0$.
In our case, one could set the sign of $\phi^b(p)$ by adding to the action a constant
external chromo-magnetic field coupled to the gluon field or, equivalently, one can
fix this sign by a global gauge transformation. This is analogous to using a global
rotation in a spin system in order to fix a direction for the magnetization.
In all cases, i.e.\ introducing an external magnetic field, making a
global transformation, or taking the absolute value of the order parameter,
one explicitly breaks the global symmetry of the
system choosing one of the (equivalent) possible vacua.
In the present work we will consider the absolute value of $\phi^b(p)$,
but we also describe (in Section \ref{sec:glob}) a possible global 
gauge transformation to select a unique broken vacuum.

\begin{figure}[b!]
\begin{center}
\protect\hskip -1.2cm
\protect\vskip 0.4cm
\includegraphics[height=0.80\hsize]{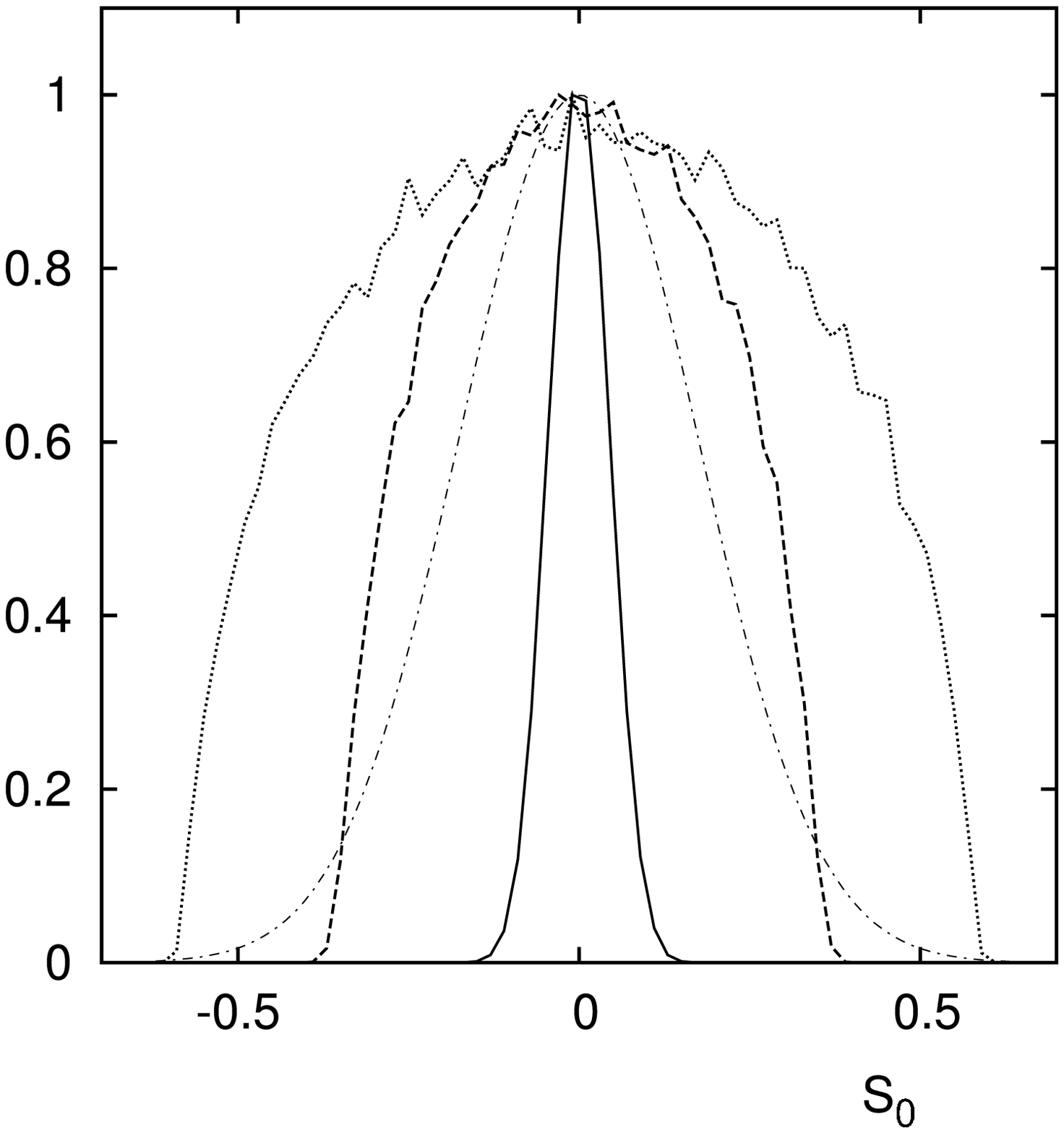}
\caption{Histograms for the first component of the magnetization in
the $3d$ $O(4)$ nonlinear $\sigma$-model. Solid, dashed and dotted lines
correspond respectively to inverse temperature $\beta = 0.9$ (in the 
symmetric phase), $\beta = 0.98$ and $\beta = 1.2$ (both in the broken 
phase). In the three cases we considered 100,000 independent configurations. 
For comparison, we show (dot-dashed line) a Gaussian with the same mean and
variance as the case $\beta = 0.98$.
\label{fig:hist}
}
\end{center}
\end{figure}

\vskip 2mm

Another indication of spontaneous symmetry breaking is the direct
investigation of the statistical distribution of the order parameter,
obtained from histograms of the produced Monte Carlo data.
Let us recall that, given a random variable $x$ with a Gaussian distribution
$ e^{-x^2 / (2 a)} / \sqrt{2 \pi a}$, corresponding to a null mean value 
and a standard deviation $\sqrt{a}$, 
one finds that the variable $| x |$ has mean value
$\,\sqrt{ 2\,a/ \pi}$ and standard deviation
$\, \sqrt{(\pi - 2) \, a / \pi} \approx 0.75 \, \langle \, | x | \, \rangle $.
Thus, if one evaluates the quantity $x$ and finds that its average is null within
error bars while $| x |$ has a relatively small error, one should conclude that
the data for $ x $ do not correspond to a Gaussian distribution and that the
fluctuations observed are not just statistical ones, since in this case they
should be large also when considering the absolute value of $ \, x $. This is
indeed what happens for the magnetization of a spin system in the broken 
(low-temperature) phase, as can be easily verified by comparing a histogram 
of the data with a Gaussian distribution. 
This is true both for discrete symmetries --- as in the Ising model, for which
the distribution of the magnetization has two peaks
\cite{Binder-Heermann,Mouritsen} ---
and for continuous symmetries, such as for the three-dimensional
$O(N)$-symmetric nonlinear $\sigma$-models.
In this case (see Fig.\ \ref{fig:hist}), the shape of the statistical
distribution of the data is similar to a Gaussian, but with a larger width
around the peak of the curve.
Moreover, the width becomes larger as the temperature of the system decreases.
On the contrary, in the unbroken phase, the histogram of the data is indeed
a Gaussian distribution (see again Fig.\ \ref{fig:hist}). Note that a simple way of
evaluating the Gaussian character of a distribution \cite{Binder-Heermann}
is to consider the so-called Binder cumulant 
\beq
U\, =\, 1 - \frac{\langle \, x^4 \, \rangle}{3 \, \langle \, x^2 \, \rangle^2}
\label{eq:binder}
\eeq
for the order parameter $x$. In the infinite-volume limit this quantity is zero
if $x$ has a Gaussian distribution, i.e.\ in the unbroken phase, and 
is equal to $2/3$ if the distribution of $x$ has two peaks, as in the broken
phase of the Ising model. Thus, in general one should expect
$U \neq 0$ in the broken phase, where the value of $\,U\,$
depends on the distribution of $x$.
In the case of the ghost condensate, based on the discussion above, we argue
that the occurrence of large fluctuations
for $ \phi^b(p) $ but not for $\, | \, \phi^b(p) \, | \,$
implies that the global symmetry $SL(2,R)$ is indeed broken.
We will analyze the statistical distribution for
$\phi^b(p) $ in Section \ref{res}.

\vskip 2mm

As mentioned above, we show in Eq.\ (\ref{eq:phimom}) that the quantity 
$\phi^b(p)$ is proportional
to the Fourier-transformed gluon field ${\widetilde A}(q)$. This dependence
can explain the $1/L^2$ behavior of $\phi^b(p)$ as induced by
${\widetilde A}(q)$. Indeed, for $q = 0$ it was 
proven \cite{Zwanziger:1990by} that 
$\langle \, | \, {\widetilde A}(0) \, | \, \rangle $ goes to zero in the
infinite-volume limit at least as fast as $1/L$.
Furthermore, in Section \ref{res} we verify from our data that
$|\, {\widetilde A}^{b}(q) \,|$ behaves as $1/L^2$.
We stress that this result does not imply that $|\phi^b(p)|$ is zero
at infinite volume, because 
$\langle\, | \, \phi^b(p) \, | \,\rangle$ is not simply proportional 
to $\langle \, |\, {\widetilde A}^{b}(q) \,| \,\rangle$,
as can be seen in Eq.\ \reff{eq:phimodul}.

\vskip 2mm

The paper is organized as follows. In Section \ref{fp} we review the
definition of the Faddeev-Popov operator --- from which we obtain the ghost
propagator --- on the lattice (for the
minimal Landau gauge) and in the continuum (in Landau gauge) for a
general $SU(N_c)$ gauge group. In Section \ref{cond} we define
$\phi^b(p)$ and show that it is proportional to the Fourier-transformed 
gluon field ${\widetilde A}(q)$. In that Section we also discuss
how to find a global gauge transformation that changes the sign of $\phi^b(p)$.
Our numerical simulations are explained in Section \ref{num} and the results
are reported in Section \ref{res}. Let us stress that, as a by-product of our analysis,
we also present a careful study of the color structure of the inverse
Faddeev-Popov matrix in momentum space.
At the same time, we consider the infinite-volume limit for the
momentum-space gluon field components ${\widetilde A}(q)$.
Finally, in Section \ref{con} we draw our conclusions.


\section{The Landau Faddeev-Popov operator}
\label{fp}

Let us recall that, on a $d$-dimensional lattice, the (minimal) Landau
gauge is obtained, in the $SU(N_c)$ case, by considering the functional
\beqa
\!\!\!\!\!\!\!\!\! {\cal E}_{U}[ g ] \!  & = & \!
        - \frac{1}{d\,V} \sum_{x} \sum_{\mu}
            \Re \, \frac{\tr}{N_c} \left[
 g(x) \, U_{\mu}(x) \, g^{\dagger}(x + e_{\mu})
                \right] \label{eq:defEori} \\[2mm]
& = & - \frac{1}{d\,V} \sum_{x} \sum_{\mu} \,
                  \frac{\tr}{2\,N_c} \,\left[
        \, g(x) \, U_{\mu}(x) \, g^{\dagger}(x + e_{\mu}) 
                     \right. \nonumber\\[2mm]
   & & \left. \quad \;\;\;\;\;\;\; + \, g(x + e_{\mu}) \, U^{\dagger}_{\mu}(x) \, g^{\dagger}(x) 
              \right]
\label{eq:defE}
\,\mbox{.}
\eeqa
Here $\{ U_{\mu}(x) \}$ is a given ({\em i.e.}\ fixed)
thermalized lattice configuration, $ V $ is the lattice volume
and the gauge-fixing
condition is imposed by finding a gauge transformation
$\{ g (x) \}$ that brings this functional to a minimum.
Both the link variables $ U_{\mu}(x) $ and the
gauge transformation matrices $ g (x) $ are elements of
the $SU(N_c)$ group (in the fundamental $N_c \times N_c$ representation).
Also, $\1$ indicates the identity matrix and $e_{\mu}$ is a
positive unit vector in the $\mu$ direction. Note that we
set the lattice spacing $a$ equal to 1 in order to simplify the
notation.

Let us also consider \cite{Zwanziger:1993dh,Cucchieri:1997ns}
a one-parameter subgroup
$g(\tau; x ) = \exp \left[ \;i\, \tau \, \gamma^b(x) \lambda^b
\; \right]$,
where $\tau$ is a real parameter, $\lambda^b$ are the $N_c^2 - 1$
traceless Hermitian generators of $SU(N_c)$ (also in the
fundamental representation), the components of $\gamma^b(x)$ are real
and the sum over the color
index $ b $ is understood. We consider generators $\lambda^b$ normalized
such that $\tr \, ( \lambda^b \lambda^c ) = 2 \, \delta^{bc}$. In the $SU(2)$
case these generators are the three Pauli matrices $\sigma^{b}$.
Using this one-parameter subgroup we can regard ${\cal E}_{U}$
as a function of the parameter $\tau$. Its
first derivative with respect to $\tau$ calculated at $\tau = 0$
is given by
\beq
{\cal E}^{'}(0) \, = \,
  \frac{2}{d\,V\,N_c} \, \sum_{x} \, \gamma^{b}(x)  \,
          \left(\nabla\cdot
        A^b \right)(x)
  \, \mbox{.}
\label{eq:Ederiv}
\eeq
Here
\beq
\left( \nabla\cdot A^b \right)(x) \,=\,
  \sum_{\mu} \, A_{\mu}^{b}(x) -
                  A_{\mu}^{b}(x - e_{\mu})
\label{eq:diverA}
\eeq
is the lattice divergence of the gluon field $A_{\mu}^{b}(x)$, defined as
\beq
A_{\mu}^{b}(x) \,=\, \frac{1}{2} \,
\tr \; [\, A_{\mu}(x)\, \lambda^{b} \, ]
\label{eq:A}
\eeq
with~\footnote{Note that in Eq.\ (\ref{eq:Atrac}) one should define
the gluon field at the midpoint $A_{\mu}(x+e_{\mu}/2)$
(see for example \cite{Leinweber:1998uu}), but in our case we may ignore this subtlety.}
\beq
A_{\mu}(x) \,=\, \frac{1}{2 \,i}\,
\left[ \, U_{\mu}(x) - U_{\mu}^{\dagger}(x) \, \right]_{\mbox{traceless}}
\label{eq:Atrac}
\eeq
or, equivalently,
\beqa
 A_{\mu}(x) & = & \frac{1}{2 \,i}\,
\left[ \, U_{\mu}(x) - U_{\mu}^{\dagger}(x) \, \right]
 \nonumber \\[2mm]
  & & \;\;\; -\,  \1\, \frac{\tr}{2 \,i\,N_c}\,
\left[ \, U_{\mu}(x) - U_{\mu}^{\dagger}(x) \, \right]
\,\mbox{.}
\label{eq:defAgen}
\eeqa
The last term in the above equation is zero in the $SU(2)$ case
and the definition of the lattice gluon field simplifies to
\beq
A_{\mu}(x) \,=\, \frac{1}{2 \,i}\,
\left[ \, U_{\mu}(x) - U_{\mu}^{\dagger}(x) \, \right] \,\mbox{.}
\label{eq:defAsym}
\eeq
Then, one can write the $SU(2)$ link variables $U_{\mu}(x)$ as
\beq
U_{\mu}(x) \;=\; U^0_{\mu}(x) \, \1\,+\,i\,A^b_{\mu}(x)\, \sigma^b
\label{eq:Udecomp}
\eeq
with
\beq
\left[U^0_{\mu}(x)\right]^2 \,+\,\sum_b \, \left[A^b_{\mu}(x)\right]^2 \,=\, 1
\,\mbox{.}
\eeq
Also, note that, since the generators
$\lambda^b$ are traceless, the term proportional to the identity
matrix $\1$ in Eq.\ (\ref{eq:defAgen})
does not contribute to $A_{\mu}^{b}(x)$ [see Eq.\ (\ref{eq:A})].
Thus, for any $N_c \geq 2$ we find
\beqa
\!\!\!\!\!\! A_{\mu}(x) &=& \lambda^b \, A_{\mu}^{b}(x)
       \;=\; \frac{\lambda^b}{2}\,\tr 
   \left[ \, A_{\mu}(x) \, \lambda^b \,\right] \\[2mm]
           &=& \frac{\lambda^b}{4\,i}\,\tr \left\{\,
   \left[ \, U_{\mu}(x) - U_{\mu}^{\dagger}(x) \, \right] \, \lambda^b \,\right\} \\[2mm]
           &=& \frac{\lambda^b}{2}\, \mbox{Im} \tr 
   \left[ \, U_{\mu}(x) \, \lambda^b \,\right]
\,\mbox{.}
\label{eq:Asimpl}
\eeqa

If $\{ U_{\mu}\left(x\right) \}$ is a stationary point of
${\cal E}(\tau)$ at $\tau = 0$
then we have
${\cal E}^{'}(0) = 0$ for every $\{ \gamma^{b}(x) \}$. This
implies [see Eq.\ (\ref{eq:Ederiv})]
\beq
\left( \nabla \cdot A^{b} \right)(x) = 0
\label{eq:diverg0}
\eeq
for any $x$ and $b$, which is the
lattice formulation of the usual Landau gauge-fixing condition
in the continuum.
Let us recall that the continuum gluon field $A^{(c)}_{\mu}(x)$~\footnote{Here we use
the superscript $(c)$ for quantities in the continuum only when necessary to avoid
confusion between these and the corresponding lattice quantities.}
is related to the link variables $U_{\mu}(x)$ by the relation
\beq
U_{\mu}(x) \;=\; \exp{\left[i\,a\,g_0\,A^{(c)}_{\mu}(x)\right]}
\, \mbox{,}
\label{eq:defAmath}
\eeq
where $g_0$ is the lattice bare coupling constant. This yields~\footnote{Note that,
in the $SU(N_c)$ case
with $N_c \geq 3$, the second term on the r.h.s.\ of
Eq.\ (\ref{eq:defAgen}) is of order $a^3$. Thus,
the definition (\ref{eq:defAsym}) can be used also for the $SU(3)$ case,
corresponding to a different lattice discretization for the
gluon field \cite{Marenzoni:1993td,Marenzoni:1994ap}.}
\beq
A_{\mu}(x) \;=\;
a\,g_0\,A^{(c)}_{\mu}(x) \;+\; {\cal O}(a^3) \,\mbox{.}
\eeq
Finally, using
\beq
\nabla_{\mu} \;=\; a\,\partial_{\mu} \,+\,{\cal O}(a^2)
\label{eq:nabla}
\eeq
we obtain from Eq.\ (\ref{eq:diverg0}) the result
\beq
\left( \partial \cdot A^b \right)(x) \,=\, {\cal O}(a) 
  \,\mbox{.}
\eeq

Still considering the one-parameter subgroup $g(\tau; x)$ and the general
$SU(N_c)$ case
we can evaluate the second derivative of ${\cal E}(\tau)$
at $\tau = 0$. We obtain
\beqa
\!\!\!\!\!\!\! {\cal E}^{''}(0) &=& \frac{2}{d\,V\,N_c} \sum_{x\mbox{,}\,y}\, \sum_{b\mbox{,}\,c}\,
   \gamma^{b}(x)\,
{\cal M}_{U}^{bc}(x\mbox{,}y)\,
\gamma^{c}(y) \label{eq:secderofE2}\\
& = & \,\frac{2}{d\,V\,N_c} \sum_{x}\,\sum_{b}\,
    \gamma^{b}(x)\, \left( {\cal M}_{U} \gamma \right)^b(x)
\,\mbox{,}
\label{eq:Mprodu}
\eeqa
with \cite{Zwanziger:1993dh}
\beqa
\left( {\cal M}_U \gamma \right)^b(x) & = & \sum_{\mu} \, \Gamma_{\mu}^{bc}(x) \left[ \, \gamma^c(x) \,-\,
                                      \gamma^c(x+e_{\mu}) \, \right] \nonumber \\
                        &  & \;\; + \, \Gamma_{\mu}^{bc}(x-e_{\mu}) \left[  \,\gamma^c(x) \,-\,
                                      \gamma^c(x-e_{\mu}) \, \right] \nonumber \\[2mm]
                  & & \;  \; \; + f^{b d c} \,
                    \left[ \, A^d_{\mu}(x) \, \gamma^c(x+e_{\mu}) \right.  \nonumber \\[2mm]
                   & & \left. \;\;\; \; \; \; - \, A^d_{\mu}(x-e_{\mu}) \, \gamma^c(x-e_{\mu}) \, \right]
\label{eq:Mdef}
\,\mbox{.}
\eeqa
Here ${\cal M}_{U}^{bc}(x\mbox{,}y)$
is the lattice Faddeev-Popov matrix and
$f^{b c d}$ are the (anti-symmetric) structure constants defined
by $\left[ \lambda^b\mbox{,}\,\lambda^c \right] = 2 \,i f^{b c d} \lambda^d$.
We also have \cite{Zwanziger:1993dh}
\beq
\Gamma_{\mu}^{bc}(x) \,=\, \tr \, \left[  
                 \left\{ \,\frac{\lambda^b}{2}\,\mbox{,}\frac{\lambda^c}{2}\,\right\} 
 \frac{U_{\mu}(x) + U^{\dagger}_{\mu}(x)}{2}
                 \right]
\label{eq:defgamma}
\,\mbox{.}
\eeq
In the $SU(2)$ case \cite{Bakeev:2003rr} one finds $\Gamma_{\mu}^{bc}(x) = \delta^{bc} \tr U_{\mu}(x) / 2$
and $f^{b c d} \! = \! \epsilon^{b c d}$, where
$\epsilon^{b c d}$ is the completely anti-symmetric tensor.
In this case, Eqs.\ (\ref{eq:Mprodu}) and (\ref{eq:Mdef}) agree with Eq.\ (11) of
Ref.\ \cite{Cucchieri:1997ns}.
Note that, by using Eqs.\ ({\ref{eq:diverA}) and (\ref{eq:diverg0}),
we can re-write Eq.\ (\ref{eq:Mdef}) as
\beqa
\!\!\!\!\!\!\!\!\!\!\!
  \left( {\cal M}_U \gamma \right)^b(x) & = & \sum_{\mu} \, \Gamma_{\mu}^{bc}(x) \left[\, \gamma^c(x) \,-\,
                                      \gamma^c(x+e_{\mu}) \,\right] \nonumber \\     
                        & + & \! \Gamma_{\mu}^{bc}(x-e_{\mu}) \left[\, \gamma^c(x) \,-\,
                                      \gamma^c(x-e_{\mu}) \,\right] \nonumber \\[2mm]
                  & + & \! f^{b d c} \left\{ A^d_{\mu}(x) \, \left[\, \gamma^c(x+e_{\mu})
                                  \,+\, \gamma^c(x) \,\right] \right.  \nonumber \\[2mm]
                   & - & \!\!\! \left. A^d_{\mu}(x-e_{\mu}) \left[\, \gamma^c(x-e_{\mu}) \,+\,
                                                         \gamma^c(x) \,\right] \right\}
\label{eq:MD}
\,\mbox{.}
\eeqa
We then obtain
\beqa
\!\!\!\!\!\!\!\!\!\! \left( {\cal M}_U \gamma \right)^b(x) &=& \sum_{\mu} \, \left( D_{\mu} \gamma \right)^b(x) \,-\,
                        \left( D_{\mu} \gamma \right)^b(x-e_{\mu}) \\[2mm]
         &=& \left(\nabla \cdot D_{\mu} \gamma \right)^b(x)
\,\mbox{,}
\label{eq:MlatticeD}
\eeqa
where we used the definition (\ref{eq:diverA}) of the lattice divergence and
\beqa
\!\!\!\!\!\! \left( D_{\mu} \gamma \right)^b(x) &=& \Gamma_{\mu}^{bc}(x) \left[\, \gamma^c(x) \,-\,
                                      \gamma^c(x+e_{\mu}) \,\right] \nonumber \\[2mm]
                    &+& \! f^{b d c} \, A^d_{\mu}(x) \, \left[\, \gamma^c(x+e_{\mu})
                                  \,+\, \gamma^c(x) \,\right]
\label{eq:defcovarD}
\eeqa
is the (negative of the) lattice covariant derivative \cite{Zwanziger:1993dh} applied to $\gamma^c(x)$.
Equivalently, we can write Eq.\ (\ref{eq:Mdef}) as
\beqa
\!\!\!\!\!\!\!\!\!\!\!\!\!
  \left( {\cal M}_U \gamma \right)^b(x) & = & \sum_{\mu} \, \Gamma_{\mu}^{bc}(x) \left[\, \gamma^c(x) \,-\,
                                      \gamma^c(x+e_{\mu}) \,\right] \nonumber \\
                        & + & \! \Gamma_{\mu}^{bc}(x-e_{\mu}) \left[\, \gamma^c(x) \,-\,
                                      \gamma^c(x-e_{\mu}) \,\right] \nonumber \\[2mm]
                  & + & \! f^{b d c} \left\{ A^d_{\mu}(x) \, \left[\, \gamma^c(x+e_{\mu})
                                  \,-\, \gamma^c(x) \,\right] \right.  \nonumber \\[2mm]
                   & - & \left. A^d_{\mu}(x-e_{\mu}) \left[\, \gamma^c(x-e_{\mu}) \,-\,
                                                         \gamma^c(x) \,\right] \right\}
\label{eq:Msym}
\eeqa
and it is evident that the second derivative  ${\cal E}^{''}(0)$
is null if the vector $\gamma^{c}(x)$
is constant, i.e.\ the Faddeev-Popov matrix has a
trivial null eigenvalue corresponding to a constant eigenvector. It is also
clear that, if $\{ U_{\mu}\left(x\right) \}$ is a local minimum of
${\cal E}(\tau)$ at $\tau = 0$, then the matrix ${\cal M}_{U}$ is
positive definite (in the subspace orthogonal to the space of
constant vectors).

Let us recall that, in the continuum, if one considers
a gauge-fixing condition $\,\left[ F(A^{g}) \right]^{b}(x) \! = \! 0$ and
the gauge transformation $g(x) = e^{i \, g_0\, \theta^b(x) \,\lambda^b}$,
the Faddeev-Popov operator can be determined using the expansion
\beqa
\!\!\!\!\!\!\!\!\! \left[ F(A^{g}) \right]^{b}(x) &=& 
    \left[ F(A) \right]^{b}(x) \nonumber \\[2mm]
     & + & \! \!\int \, d^4y\,
          \sum_c \, {\cal M}^{bc}(x,y) \, \theta^c(y) \;+\; \ldots
\;\mbox{.}
\eeqa
In the Landau case one has $ \left[ F(A) \right]^{b}(x)
= (\partial_{\mu} \, A_{\mu}^b)(x)$, giving
\beqa
\!\!\!\!\!\!\!\!\!\!\!\! {\cal M}^{bc}(x,y) \!\! &=& \!\! - \partial_{\mu}\, D^{bc}_{\mu}(x) \, \delta(x - y) 
\label{eq:MofD} \\[2mm]
                                          &=& \!\! - D^{bc}_{\mu}(x) \, \partial_{\mu}\, \delta(x - y) \\[2mm]
            &=& \!\! - \left[ \delta^{bc}\,\partial_{\mu} - 2\, g_0\,
                          f^{bdc} A^{d}_{\mu}(x) \right]\partial_{\mu}\, \delta(x - y)
\label{eq:Mcont1} \\[2mm]
            &=& \!\! - \left[ \delta^{bc}\,\partial_{\mu} + 2\, g_0\,
                          f^{bcd} A^{d}_{\mu}(x) \right]\partial_{\mu}\, \delta(x - y)
         \, \mbox{,}
\label{eq:Mcont}
\eeqa
where $D^{bc}_{\mu}(x)$ is the (continuum) covariant derivative.
Note that, with our normalization for the generators $\lambda^b$, we have
that $2\,A^{d}_{\mu}(x)$ is the standard continuum gluon field.

Clearly, we can expand Eq.\ (\ref{eq:defAmath}) in the limit $a \to 0$
as
\beq
U_{\mu}^{0}(x) \;=\; \1 \,+\,i\,a\,g_0\,A_{\mu}(x) \,+\, {\cal O}(a^2)
\,\mbox{,}
\eeq
yielding
\beq
\Gamma_{\mu}^{bc}(x) \;\approx\; \delta^{bc} + {\cal O}(a^2) \, \mbox{.}
\eeq
Also, after Taylor expanding the vectors $\gamma^b(x \pm e_{\mu}) = \gamma^b(x \pm a e_{\mu})$
to order $a^2$ in Eq. (\ref{eq:Mdef}), one obtains
\beq
a^{-2} \, {\cal M}_{U}^{bc}(x\mbox{,}y) \;=\; {\cal M}^{bc}(x,y)
 \,+\,{\cal O}(a) \, \mbox{,}
\eeq
with ${\cal M}^{bc}(x,y)$ given in Eq.\ (\ref{eq:Mcont1}).
At the same time, from Eq.\ (\ref{eq:defcovarD}) we find the relation between
the lattice and the continuum expressions for the covariant derivative
\beq
D^{bc}_{\mu}(x) \;=\; -  a \, D^{(c) \, bc}_{\mu}(x) \,+\, {\cal O}(a^2)
\eeq
and it is immediate to obtain Eq.\ (\ref{eq:MofD}), using Eqs.\ (\ref{eq:MlatticeD}) and (\ref{eq:nabla}).


\section{The ghost condensate}
\label{cond}

Let us note that the lattice Faddeev-Popov matrix
${\cal M}_{U}^{bc}(x\mbox{,}y) $ is symmetric with respect
to the exchange $(b, x) \leftrightarrow (c, y)$. Indeed, from Eq.\ (\ref{eq:Mdef})
we can write \cite{Sternbeck:2005tk}
\beqa
\!\!\!\!\!\!\!\!\!\!\!
  {\cal M}_U^{bc}(x,y) & = & \sum_{\mu} \, \Gamma_{\mu}^{bc}(x) \left[\, \delta_{x,y} \,-\,
                                      \delta_{x+e_{\mu},y} \,\right] \nonumber \\
                        & + & \Gamma_{\mu}^{bc}(x-e_{\mu}) \left[\, \delta_{x,y} \,-\,
                                      \delta_{x-e_{\mu},y} \,\right] \nonumber \\[2mm]
                  & + & f^{b d c} \left[ A^d_{\mu}(x) \, \delta_{x+e_{\mu},y} \right. \nonumber \\[2mm]
                   &  & \left. \;\;\;\; -\, A^d_{\mu}(x-e_{\mu}) \delta_{x-e_{\mu},y} \, \right]
\label{eq:Melements}
\eeqa
and it is straightforward to verify that ${\cal M}_U^{bc}(x,y) =
{\cal M}_U^{cb}(y,x)$. Moreover, using the fact that the inverse of a
symmetric matrix is also symmetric, we have that the same property is
satisfied by the inverse matrix
\beq
G^{bc}(x\mbox{,}y) = ({\cal M}_U^{-1})^{bc}(x\mbox{,}y)
\,\mbox{.}
\label{eq:Ginv}
\eeq
Let us recall that, if a matrix $M^{bc}(x,y)$
has two different sets of indices, it can always be decomposed as
\beqa
M^{bc}(x,y) &=& M_{ss}^{bc}(x,y) \,+\, M_{sa}^{bc}(x,y) \nonumber \\[2mm]
            & & \;\;\; +\ M_{as}^{bc}(x,y) \,+\, M_{aa}^{bc}(x,y)
\,\mbox{,}
\eeqa
where the sub-scripts $s$ and $a$ indicate symmetry and anti-symmetry
with respect to the color or to the space-time indices (in this order).
Furthermore, if the matrix $M^{bc}(x,y)$ is symmetric with respect
to the exchange $(b, x) \leftrightarrow (c, y)$, as in the case of the Landau 
Faddeev-Popov matrix ${\cal M}_U$, then this decomposition simplifies to
\beq
M^{bc}(x,y) \;=\; M_{ss}^{bc}(x,y) \,+\, M_{aa}^{bc}(x,y)
\label{eq:decsym}
\,\mbox{.}
\eeq
Indeed, for the Faddeev-Popov matrix ${\cal M}_{U}^{bc}(x\mbox{,}y) $ 
[see Eq.\ (\ref{eq:Melements})] we have that
the terms containing $\Gamma^{bc}_{\mu}(x)$ are symmetric under
the exchange $b \leftrightarrow c$ and under the
exchange $x \leftrightarrow y$. At the same time, the terms
containing $f^{bcd}$ are anti-symmetric in the color indices
and in the space-time coordinates $x$ and $y$. Thus, in the
$SU(2)$ case,
we can write [see Eqs.\ (\ref{eq:Mdef}) and (\ref{eq:defgamma})]
\beq
{\cal M}_{U}^{bc}(x\mbox{,}y) \;=\; \delta^{b c} {\cal S}(x\mbox{,}y) \,-\,
                                         f^{b c d} {\cal A}^d(x\mbox{,}y)
\label{eq:Mdeco}
\eeq
with ${\cal S}(x\mbox{,}y) = {\cal S}(y\mbox{,}x)$
and ${\cal A}^d(x\mbox{,}y) = - {\cal A}^d(y\mbox{,}x)$, corresponding to the
decomposition (\ref{eq:decsym}). Note that the above formula also holds
in the continuum [see Eq.\ (\ref{eq:Mcont})] for any $N_c \geq 2$.
Then, the inverse Faddeev-Popov matrix is given by
\beqa
\!\!\!\!\!\!\!\!\!\!
\left({\cal M}_{U}^{-1} \right)^{bc}(x\mbox{,}y) &=& \left\{ \delta^{b e} {\cal S}(x\mbox{,}w)
        \left[ \delta^{e c} \delta_{w,y} \right. \right. \nonumber \\[2mm]
  & & \; - \, \left. \left. f^{e c d} 
    {\cal S}^{-1}(w\mbox{,}z) {\cal A}^d(z\mbox{,}y) \right] \right\}^{-1}
 \,\mbox{,}
\eeqa
where the sum over repeated indices is understood. This implies
that the inverse matrix $G^{bc}(x\mbox{,}y)$ defined in Eq.\ \reff{eq:Ginv}
above can be evaluated
using~\footnote{This expansion can be used to evaluate numerically
the ghost propagator (see for example \cite{Furui:2003jr}).}
\beqa
G^{bc}(x\mbox{,}y) &=& \delta^{b c} {\cal S}^{-1}(x\mbox{,}y) \nonumber\\[2mm]
                     &+&
       f^{b c d} {\cal S}^{-1}(x\mbox{,}z) {\cal A}^d(z\mbox{,}w) {\cal S}^{-1}(w\mbox{,}y)
                \nonumber\\[2mm]
     &+& \left[ \,f^{b d h} f^{d c j} {\cal S}^{-1}(x\mbox{,}z) {\cal A}^h(z\mbox{,}s)
                    {\cal S}^{-1}(s\mbox{,}r) \right. \nonumber \\[2mm]
           & & \left. \;\;\; {\cal A}^j(r\mbox{,}w) {\cal S}^{-1}(w\mbox{,}y) \,\right] \,+\, \ldots
\;\mbox{.}
\label{eq:Gexp}
\eeqa
We can also write
\beq
G^{bc}(x\mbox{,}y) \;=\; G_e^{bc}(x\mbox{,}y) \,+\, G_o^{bc}(x\mbox{,}y)
\,\mbox{.}
\eeq
Here
\beqa
\!\!\!\!\!\!
  G_e^{bc}(x\mbox{,}y) &=& \delta^{bc} {\cal S}^{-1}(x\mbox{,}y) \,+\,
      f^{b d h} f^{d c j} \, \left[ \right. \nonumber \\[2mm]
           & & \left. \;\;\; {\cal S}^{-1} \, {\cal A}^h \, {\cal S}^{-1} \,
                  {\cal A}^j \, {\cal S}^{-1} \,\right](x\mbox{,}y) \,+\, \ldots
\label{eq:Gpropexp}
\eeqa
includes all terms of Eq.\ (\ref{eq:Gexp}) with an even number of factors
${\cal A}^d(z\mbox{,}w)$ and
\beqa
\!\!\!\!\!
G_o^{bc}(x\mbox{,}y) \!\!&=&\!\! f^{b c d} \,\left[ \,{\cal S}^{-1} \, {\cal A}^d \,
                                {\cal S}^{-1} \,\right](x\mbox{,}y) \,+\, f^{b d h} f^{d e j} f^{e c l} \, \left[
                                             \right. \nonumber \\[2mm]
    & &\!\! \left. {\cal S}^{-1} \, {\cal A}^h \,
                                          {\cal S}^{-1} \,
                       {\cal A}^j \, {\cal S}^{-1} \,
                             {\cal A}^l \, {\cal S}^{-1} \,\right](x\mbox{,}y) \,+\, \ldots
\label{eq:Gcondexp}
\eeqa
is obtained by considering
all terms with an odd number of factors ${\cal A}^d(z\mbox{,}w)$.
Then, it is easy to verify that
\beqa
G_{ss}^{bc}(x\mbox{,}y) &=& \frac{1}{2} \, \left[ \, G^{bc}(x\mbox{,}y) \,+\,G^{cb}(y\mbox{,}x)\,\right] \\[2mm]
                           &=& G_{e}^{bc}(x\mbox{,}y) \\[2mm]
G_{aa}^{bc}(x\mbox{,}y) &=& \frac{1}{2} \, \left[ \, G^{bc}(x\mbox{,}y) \,-\,G^{cb}(y\mbox{,}x)\,\right] \\[2mm]
                           &=& G_{o}^{bc}(x\mbox{,}y)
\,\mbox{,}
\eeqa
[where we considered again the decomposition (\ref{eq:decsym})], and
that~\footnote{Note that Eq.\ (\ref{eq:Gfr}) agrees with Eq.\ (20)
in Ref.\ \cite{Boucaud:2005ce}.}
\beqa
G_o^{bc}(x\mbox{,}y) &=& f^{b d e} \, {\cal S}^{-1}(x\mbox{,}z)\,
                                   {\cal A}^e(z\mbox{,}w)\,G_e^{dc}(w\mbox{,}y) \\[2mm]
G_e^{bc}(x\mbox{,}y) &=& \delta^{bc} {\cal S}^{-1}(x\mbox{,}y) \nonumber\\[2mm]
     &+& f^{b d h} \, {\cal S}^{-1}(x\mbox{,}z)\, {\cal A}^h(z\mbox{,}w)\, G_o^{dc}(w\mbox{,}y)
\,\mbox{.}
\label{eq:Gfr}
\eeqa
Thus, for the ghost propagator $G(x\mbox{,}y)$ --- defined on the lattice as 
$\tr {\cal M}_{U}^{-1} / (N_c^2 - 1) $ --- we obtain
\beq
G(x\mbox{,}y) \;=\; \frac{\delta^{b c}}{N_c^2 - 1} \,G^{bc}(x\mbox{,}y) \;=\;
          \frac{\delta^{b c}}{N_c^2 - 1} \,G_e^{bc}(x\mbox{,}y) \,\mbox{,}
\label{gp0}
\eeq
while for ghost condensation (in the Overhauser channel) we consider 
\cite{Dudal:2003dp} the quantity
\beqa
\!\!\!\!\!\!\!\!\!\!
\phi^b(x\mbox{,}y) &=& \frac{1}{N_c} f^{bcd}\, G^{cd}(x\mbox{,}y) \;=\;
                   \frac{1}{N_c} f^{bcd}\, G_o^{cd}(x\mbox{,}y) \\[2mm]
     &=& \frac{f^{bcd}\, f^{c e h}}{N_c} \, {\cal S}^{-1}(x\mbox{,}z)\,
                                   {\cal A}^h(z\mbox{,}w)\,G_e^{ed}(w\mbox{,}y)
\,\mbox{.}  
\label{eq:phiasG}
\eeqa
Also, in momentum space we have
\beqa
\!\!\!\!\!\!\!\!\!\! G(p)
      \!\!&=&\!\! \frac{\delta^{bc}}{V \left(N_c^2 - 1\right)} \, \sum_{x\, y} \,
                G_e^{bc}(x\mbox{,}y) \,
            e^{- \,i \, p\, (x - y)} \\[2mm]
     \!\!&=&\!\! \frac{\delta^{bc}}{V \left(N_c^2 - 1\right)} \, \sum_{x\, y} \,
             G_e^{bc}(x\mbox{,}y) \,
            \cos{\left[\,p\, (x - y)\,\right]}
\, \mbox{,}
\eeqa
where in the last equation
we used the symmetry of $G_e^{bc}(x\mbox{,}y)$
with respect to the space-time coordinates $x$ and $y$. At the same time, we find 
\beqa
\!\!\!\!\!\!\!\!\! \phi^b(p) 
     &=& \frac{f^{bcd}}{V \, N_c} \, \sum_{x\, y} \, G_o^{cd}(x\mbox{,}y) \,
            e^{- \,i \, p\, (x - y)} \\[2mm]
     &=& - i \, \frac{f^{bcd}}{V \, N_c} \, \sum_{x\, y} \, G_o^{cd}(x\mbox{,}y) \,
            \sin{\left[\, p\, (x - y) \,\right]}
\, \mbox{.}
\eeqa

Considering the result (\ref{eq:phiasG}) we can also write
\beq
\phi^b(p) \;=\; \frac{f^{bcd}\, f^{c e h} }{N_c} \, \< \, p \, | \, {\cal S}^{-1} \, | \, q \, \> \,
                                   \< \, q \, | \, {\cal A}^h \, | \, k \, \>\, 
                                   \< \, k \, | \, G_e^{ed} \, | \, p \, \>
\, \mbox{,}
\label{eq:phib}
\eeq
where we used Dirac notation and sums over $q$ and $k$ are understood.
In order to simplify our analysis we can work in the continuum. Then,
the matrix ${\cal S}(x,y)$ is given by $ - \partial^2_x \, \delta(x - y) $ and
${\cal A}^h = 2 \, g_0 \, A^{h}_{\mu} \, \partial_{\mu} \, \delta(x - y)$
[see Eq.\ (\ref{eq:Mcont})]. Thus, we obtain
\beq
\< \, p \, | \, {\cal S}^{-1} \, | \, q \, \> \;=\; \delta(p - q) \, p^{-2}
\label{eq:Sinv}
\eeq
and
\beq
\< q \, | \, {\cal A}^h \, | \, k \, \> \;=\; 2 \, i \, g_0 \, {\widetilde A}^{h}_{\mu}(q-k) \, k_{\mu}
\label{eq:calAmom}
\,\mbox{.}
\eeq
Here ${\widetilde A}^h_{\mu}(p)$ is the Fourier-transformed gluon field
defined (in the continuum and in $d$ dimensions) as
\beq
{\widetilde A}^h_{\mu}(p) \;=\; \frac{1}{\left(2 \, \pi \right)^d} \, 
      \sum_x \, A^h_{\mu}(x) \, e^{-i \, p \, x}
\,\mbox{.}
\label{eq:Ak}
\eeq
Using the above results and 
$\< \, k \, | \, G_e^{ed} \, | \, p \, \> = G_e^{ed}(k, p)$ we find
\beq
\phi^b(p) \;=\;
 \frac{f^{bcd}\, f^{c e h} }{N_c} \, \frac{2\, i\, g_0 \, k_{\mu} \,
                    \, {\widetilde A}^{h}_{\mu}(p - k) \, G_e^{ed}(k,p)}{p^2}
\, \mbox{,}
\label{eq:phimom}
\eeq
where a sum over the momenta $k$ is understood. Note that,
if $G_e^{bc}(k\mbox{,}p)$ is diagonal in color space, i.e.\ if
\beq
G_e^{bc}(k,p) = \delta^{bc} \, G_e^{bb}(k,p)
\,\mbox{,}
\eeq
Eq.\ (\ref{eq:phimom}) simplifies to
\beqa
\phi^b(p) &=& \frac{f^{bce}\, f^{c e h}}{N_c} \, \frac{2 \, i\, g_0 \, k_{\mu} \,
                     \, {\widetilde A}^{h}_{\mu}(p - k) \, G_e^{dd}(k,p)}{p^2} \\[2mm]
    &=& \frac{2 \, i\, g_0 \, k_{\mu} \,
                     \, {\widetilde A}^{b}_{\mu}(p - k) \, G_e^{dd}(k,p)}{p^2}
             \,\mbox{,}
\label{eq:phibfin}
\eeqa
where we used $f^{bce}\, f^{c e h} = N_c \delta^{bh}$.

Finally, from Eq.\ (\ref{eq:Gfr}) one obtains that the off-diagonal
{\em symmetric} components of the ghost propagator $G^{bc}_e(x\mbox{,}y)$ are proportional to
the off-diagonal {\em anti-symmetric} components $G^{bc}_o(x\mbox{,}y)$, being
\beq
G_e^{bc}(x\mbox{,}y) \;=\; f^{b d h} \, {\cal S}^{-1}(x\mbox{,}z)\,
  {\cal A}^h(z\mbox{,}w)\, G_o^{dc}(w\mbox{,}y)
\, \mbox{.}
\label{eq:Gsymmetric}
\eeq
In analogy with Eq.\ (\ref{eq:phimom}), one finds (in momentum space and in
the continuum)
\beq
G_e^{bc}(p) \;=\; f^{b d h} \, \frac{2\, i\, g_0 \, k_{\mu} \,
                    \, {\widetilde A}^{h}_{\mu}(p - k) \, G_o^{dc}(k,p)}{p^2}
\, \mbox{.}
\label{eq:GeofGo}
\eeq 
At the same time, for the ghost propagator, we can write
[using Eqs.\ (\ref{eq:Gfr}) and (\ref{gp0})]
\beqa
\!\!\!\!\!\!\!\!\!\! G(x\mbox{,}y) &=& {\cal S}^{-1}(x\mbox{,}y) \nonumber\\[2mm]
     & & \;\; + \, \frac{f^{c d h}}{N_c^2 - 1}
        \, {\cal S}^{-1}(x\mbox{,}z)\, {\cal A}^h(z\mbox{,}w)\, G_o^{dc}(w\mbox{,}y) \\[2mm]
  &=& {\cal S}^{-1}(x\mbox{,}z) \,  \Bigl[ \, \delta(z-y) \nonumber \\[2mm]
    & &   \;\; - \, {\cal A}^h(z\mbox{,}w)\, \frac{N_c}{N_c^2 - 1} \, \phi^h(w\mbox{,}y)\, \Bigr]
\label{eq:ghostpropandphi}
\,\mbox{.}
\eeqa
Note that the above expression can be used to derive
Eq.\ (27) of Ref.\ \cite{Boucaud:2005ce}.


\subsection{Global gauge transformation}
\label{sec:glob}

Let us note that the term ${\cal A}^b(x\mbox{,}y)$
is linear in the gluon field $A^d_{\mu}(x)$ [see Eqs.\ (\ref{eq:Mcont1}) and
(\ref{eq:Mdeco})]. Therefore, if we apply the
transformation
\beq
A^d_{\mu}(x) \to - A^d_{\mu}(x)
\,\mbox{,}
\label{eq:AtomA}
\eeq
then the ghost propagator $G(x\mbox{,}y)$ does not change, while the quantity
$\phi^b(x\mbox{,}y)$ gets multiplied by $-1$.
Moreover, as we have seen in the previous section,
when $G_e^{bc}(x\mbox{,}y)$ is diagonal in color space,
the quantity $\phi^b(p)$ (in the continuum) is proportional to the
Fourier-transformed gluon field ${\widetilde A}^{b}_{\mu}$
[see Eq.\ (\ref{eq:phibfin})].

Let us also stress that, for a given color index $b$,
one can always find a global gauge
transformation that changes the sign of ${\widetilde A}^b_{\mu}(q)$. For example,
in the $SU(2)$ case and with $b = 3$ this can be achieved by using
the $x$-independent gauge transformation 
\beq
g = i \,\sigma^1 \, g_1 \,+\,i \,\sigma^2 \, g_2
\label{eq:gsign}
\eeq
with $g_1^2 + g_2^2 = 1$. Indeed, if $g = i \,\sum_b \,\sigma^b \, g_b$, then
the quantity
\beqa
\!\!\!\!\!\!\!\! & & \!\!\!\! \frac{1}{V} \, g\, \left[ \, \sum_x U_{\mu}(x) \,e^{-\,i\,\widetilde{q}\,x}
     \,\right] \, g^{\dagger} \\[2mm]
& & \qquad \;\;\; \;=\; g \, \left[\, {\widetilde U}^0_{\mu}(\hat{q}) \,\1
 \,+\, i \, \sigma^c\, {\widetilde A}^c_{\mu}(\hat{q}) \, \right] \, g^{\dagger}
\label{eq:global0}
\eeqa
becomes
\beq
{\widetilde U}^0_{\mu}(\hat{q}) \,\1 \,+\,
    2\,i \, {\widetilde A}^c_{\mu}(\hat{q}) \, g_c \; \sigma^b\,g_b \,-\,
       i\, {\widetilde A}^c_{\mu}(\hat{q}) \, \sigma^c
\,\mbox{.}
\label{eq:global}
\eeq
Thus, if $g_3 = 0$ one immediately finds that
the sign of ${\widetilde A}^3_{\mu}(\hat{q})$ has changed 
(for any value of the index $\mu$ and any momentum $\hat{q}$).
Here we considered the decomposition (\ref{eq:Udecomp})
for the $SU(2)$ link variables $U_{\mu}(x)$
and the definitions
\beqa
{\widetilde U}^0_{\mu}(\hat{q}) &=& \frac{1}{V} \, \sum_x \, U^0_{\mu}(x) \,
                   e^{-\,i\,\widetilde{q}\,x} \\[2mm]
{\widetilde A}^b_{\mu}(\hat{q}) &=& \frac{1}{V} \, \sum_x \, A^b_{\mu}(x) \,
                   e^{-\,i\,\widetilde{q}\,x}
\,\mbox{.}
\label{eq:Ak0}
\eeqa

In the general $SU(N_c)$ case (with $N_c \geq 2$), changing
the sign of ${\widetilde A}^b_{\mu}(\hat{q})$ gives the relation
\beqa
\!\!\!\!\!\!\!\!\! &  & \mbox{Im} \, \tr  \,\left[ \, g \,
   \sum_{x} \, U_{\mu}(x)  \, e^{-\,i\,\widetilde{q}\,x}\, g^{\dagger}\, \lambda^b \,\right]
       \,
       \nonumber\\[2mm]
\!\!\!\!\!\!\!\!\! & & \;\;\;\;\;\;\;
 \;=\;
- \mbox{Im} \, \tr \left[\,
   \sum_{x} \, U_{\mu}(x)  \, e^{-\,i\,\widetilde{q}\,x}\, \lambda^b \,\right]
\eeqa
or
\beq
   \mbox{Im} \, \tr \, \left[ \, \sum_{x} \, U_{\mu}(x) \,
              e^{-\,i\,\widetilde{q}\,x} \, 
  \left( \,g^{\dagger}\, \lambda^b \, g\,+\,\lambda^b\,\right) \, \right] \,=\, 0
\,\mbox{,}
\eeq
where $g$ is a global gauge transformation and we used Eq.\ (\ref{eq:Asimpl}).
Then, a sufficient condition to satisfy the above equation is given by
\beq
g^{\dagger}\, \lambda^b \, g\,+\,\lambda^b \;=\;0 \,\mbox{,}
\eeq
implying
\beq
\left\{ \, \lambda^b \mbox{,} \, g \, \right\} \;=\;0
\,\mbox{.}
\eeq
Clearly, in the $SU(2)$ case with $b = 3$ this yields $g = i \,\sigma^b \, g_b$ and $g_3 = 0$.
Note that the choice
\beq
g\;=\; i\,\frac{\sigma_1\,{\widetilde A}^1_{\mu}(\hat{q})
   \,+\,\sigma_2\,{\widetilde A}^2_{\mu}(\hat{q})}{
          \sqrt{\, \left[{\widetilde A}^1_{\mu}(\hat{q})\right]^2
   \,+\, \left[{\widetilde A}^2_{\mu}(\hat{q})
\right]^2}}
\eeq
does not to change the values of
${\widetilde A}^1_{\mu}(\hat{q})$ and ${\widetilde A}^2_{\mu}(\hat{q})$.
Of course, any other choice for $g_1$ and $g_2$ (with $g_3 = 0$) would still
change the sign of ${\widetilde A}^3_{\mu}(\hat{q})$ but would modify the values of the
other two color components of ${\widetilde A}_{\mu}(\hat{q})$. In any case, a global
gauge transformation does not change the value of
${\widetilde U}^0_{\mu}(\hat{q})$ in Eq.\ (\ref{eq:global0}) above, implying that
the quantity $\sum_c  \left[ {\widetilde A}^c_{\mu}(\hat{q}) \right]^2 $  is also
left unchanged.


\section{Lattice Simulations}
\label{num}

We performed numerical simulations of pure $SU(2)$ Yang-Mills theory in 
Landau gauge considering the standard Wilson action in four dimensions~\footnote{Simulations
have been performed on several 866 MHz Pentium III and 1.7 GHz Pentium IV machines (with
256 MB RAM) at the IFSC--USP in S\~ao Carlos. The total CPU time
was about 264 days on a Pentium III machine
plus 61 days on a Pentium IV machine.}. To thermalize the 
field configurations we used the heat-bath algorithm accelerated by
{\it hybrid overrelaxation} (for details see \cite{Cucchieri:2003zx,Bloch:2003sk,
Cucchieri:2004sq}).
In Table \ref{table:data} we show the lattice volumes used in
the simulations and the corresponding numbers of
configurations. For {\em each} volume we
have considered three values of the lattice
coupling: $\beta = 2.2, 2.3$ and 2.4. The corresponding
string tensions in lattice units \cite{Bloch:2003sk,Fingberg:1992ju} are given in Table
\ref{table:string}. In the same table we also report the
lattice spacing in physical units for each coupling $\beta$,
using $\sqrt{\sigma} = 0.44 \, \mbox{GeV}$ as physical input.

\begin{table}
\begin{center}
\begin{tabular}{|c|c|c|c|c|c|}
\hline 
Lattice Volume $\,V$ & $ 8^{4} $ & $ 12^{4} $ & $ 16^{4} $ & $ 20^4$ & $ 24^{4} $   \\ \hline
No.\ of Configurations & 1000 & 700 & 400 & 300 & 100 \\ \hline
\end{tabular}
\caption{Lattice volumes and total number of configurations considered.
\label{table:data}}
\end{center}
\end{table}

\begin{table}
\begin{center}
\begin{tabular}{|c|c|c|c|}
\hline
$\beta$        & $2.2$      & $2.3$      & $2.4$   \\ \hline
$\sigma\, a^2$ & $0.220(9)$ & $0.136(2)$ & $0.071(1)$ \\ \hline
$a \, (\mbox{GeV}^{-1})$ & $1.07(2)$ & $0.838(6)$ & $0.606(4)$ \\ \hline
\end{tabular}
\caption{Lattice string tension and lattice spacing (in physical units)
for each lattice coupling $\beta$ considered in our simulations.
We used $\sqrt{\sigma} = 0.44 \, \mbox{GeV}$ as physical input.
\label{table:string}}
\end{center}
\end{table}

The minimal (lattice) Landau gauge is implemented using the
stochastic overrelaxation algorithm
\cite{Cucchieri:1995pn,Cucchieri:1996jm,Cucchieri:2003fb}.
We stop the gauge fixing when the quantity
\beq
\left(\nabla \cdot A\right)^{2} \;=\; \frac{1}{V} \,
\sum_{x} \, \sum_{b}\, \left[\, \left(\nabla\cdot
        A^b \right)(x) \right]^2
\eeq
is smaller than $10^{- 13}$.
In order to check for possible Gribov-copy effects
we have also done (for each thermalized configuration)
a second gauge fixing using the so-called smearing method 
\cite{Hetrick:1997yy}. (See again \cite{Cucchieri:2004sq} for details.)

Ghost condensation in the Overhauser channel is studied
by evaluating (here $N_c = 2$)
\beq
 \langle\, i\, \phi^b(\hat{p})\,\rangle
     \;=\; \, \frac{\epsilon^{bcd}}{2\,V} \, \sum_{x\, y} \, \langle \, G^{cd}(x\mbox{,}y)
                    \,\rangle \,
            \sin{\left[\, \hat{p}\, (x - y) \,\right]}
\label{eq:iphi}
\eeq
as a function of the lattice momentum $\hat{p}$,
with $G^{cd}(x\mbox{,}y)$ defined in Eq.\ (\ref{eq:Ginv}).
For the inversion of the lattice Faddeev-Popov matrix ${\cal M}_{U}$ we employ a
conjugate gradient (CG) method with even/odd preconditioning. Since this
inversion has to be performed in the sub-space orthogonal to the
constant modes --- corresponding to null eigenvalues of ${\cal M}_{U}$ ---
we verify at each CG iteration if the solution has indeed a zero constant
component.

We also consider the average of the absolute value of
$\phi^b(\hat{p})$. Using the continuum result (\ref{eq:phibfin}),
we should obtain
\beq
\langle\, \left| \, \phi^b(p) \,\right| \,\rangle  \;\approx\;
     \frac{6 \, g_0 \, k_{\mu} \,
                     \langle \, \left| \, {\widetilde A}^{b}_{\mu}(p - k)
          \,\right| \, G_e^{dd}(k,p) \,\rangle}{p^2}
             \,\mbox{.}
\label{eq:phimodul}
\eeq
This result is valid when the propagator $G^{ed}_e(k,p)$ is diagonal in color space, but
it should give a good approximation for $\langle\, \left| \, \phi^b(p) \,\right| \,\rangle$
also if $G^{ed}_e(k,p)$ is strongly diagonally-dominant. Let us recall
that a matrix
$M_{ij}$ is strongly (or strictly) diagonally-dominant~\footnote{See for example
Section 3.4.10 in Ref.\ \cite{Golub}.} if $M_{ii} > \sum_{j \neq i} M_{ij}$
for any $i$.
We will verify this hypothesis in Section \ref{res}.

Notice that a generic lattice
momentum $\hat{p}$ has components (in lattice units) $\hat{p}_{\mu} = 2 \,
 \sin{\left(\pi\, \widetilde{p}_{\mu} \, a / L_{\mu}\right)}$
and that the magnitude (squared) of the lattice momentum is given by
$\hat{p}^2 = \sum_{\mu}\,\hat{p}_{\mu}^2$.
Here $L_{\mu} = a \, N_{\mu}$ is the
physical size of the lattice in the $\mu$ direction,
the quantity $\widetilde{p}_{\mu}$
takes values $\lfloor \, \frac{- N_{\mu}}{2} \,\rfloor +
1\, , \ldots , \lfloor \, \frac{N_{\mu}}{2} \, \rfloor \,$ and
$N_{\mu}$ is the number of lattice points in the $\mu$ direction.
Thus, if we keep the physical size $L_{\mu} $ constant and
indicate with $p_{\mu}$ the momentum components in the continuum, we
find that the lattice components
$ \hat{p}_{\mu} = a \, p_{\mu} $ take values in the
interval $( -\pi , \, \pi ]$ when $ a \to 0 $.

In order to analyze the effects due to the breaking of rotational symmetry
we consider two types of lattice momenta, i.e.
\begin{itemize}
\item asymmetric: $\;\;\;\hat{p}_1 = \hat{p}_2 = \hat{p}_3 = 0\;$ and $\;\hat{p}_4 \neq 0\,$,
\item symmetric: $\;\;\;\hat{p}_1 = \hat{p}_2 = \hat{p}_3 = \hat{p}_4 \neq 0\,$. 
\end{itemize}

\noindent
For a symmetric lattice ($L_{\mu} = L$) we obtain
\beq
\hat{p} \;=\; \hat{p}_4 \;=\;
     2 \, \sin{\left(\frac{\pi\, \widetilde{p} \, a}{L}\right)}
\eeq
in the first case and
\beq
\hat{p} \;=\; 2\,\hat{p}_{\mu} \;=\;
     4 \, \sin{\left(\frac{\pi\, \widetilde{p} \, a}{L}\right)}
\eeq
in the second one. In both cases $\,\widetilde{p} \,$ takes the values
$ \, \frac{- N}{2} \,+
1\, , \ldots , \, \frac{N}{2} \, $ since $\, N_{\mu} = N\,$
and $\,N\,$ is always even.

Let us note that the term contributing to $\,|\phi^b(p)|\,$ when $k = p$ is given by
\beq
\frac{6 \, g_0 \, p_{\mu} \,
                     \, {\widetilde A}^{b}_{\mu}(0) \, G(p)}{p^2}
\,\mbox{.}
\label{eq:peqk}
\eeq
Thus, if we neglect all the other terms
appearing in Eq.\ (\ref{eq:phimodul}) we find
\beqa
\!\!\!\!\! \langle\, \left| \, \phi^{b}(p) \,\right| \,\rangle \! &\propto& \!
  \frac{p_{\mu} \, \langle \, \left| \, {\widetilde A}^b_{\mu}(0) \,\right| \,G(p) \,\rangle}{p^2}
\label{eq:phicontA} \\[2mm]
  \! & \sim & \! \frac{\langle \, G(p) \,\rangle}{p}
             \,\mbox{.}
\label{eq:phiGp}
\eeqa
Therefore, if the ghost propagator has the infrared behavior $\langle\,G(p) \,\rangle
\sim 1/p^{2+2\kappa}$ we obtain (in this approximation)
\beq
\langle\, \left| \, \phi(p) \,\right| \,\rangle \,\sim\, \frac{1}{p^{3+2\kappa}}
\,\mbox{.}
\eeq
Clearly, for $\kappa = 0.5$ one gets $\langle\,G(p) \,\rangle
\sim 1/p^3$ and $\langle\, \left| \, \phi(p) \,\right| \,\rangle \sim 1/p^4$.
Let us recall that for the infrared exponent $\kappa$
--- in Landau gauge and in four dimensions ---
there are several analytic predictions (using Dyson-Schwinger equations)
\cite{vonSmekal:1997is,Atkinson:1997tu,Atkinson:1998zc,Lerche:2002ep,
Fischer:2002eq,Zwanziger:2002ia,Zwanziger:2001kw}
with values in the interval $[0.52, 1.00]$.
On the other hand, numerical studies have found a value of $\kappa \approx 0.25 - 0.35$
\cite{Furui:2003jr,Sternbeck:2005tk,Cucchieri:1997dx,Bloch:2002we,Furui:2004cx},
both for the $SU(2)$ and
$SU(3)$ cases. Recently, a value $\kappa = 0.22(5)$ has been obtained in the unquenched
$SU(3)$ case \cite{Furui:2005bu}. Note that these numerical results for the infrared exponent
$\kappa$ represent a lower bound, since the value of $\kappa$ increases when one is
able to get data at smaller momenta using larger lattice volumes
\cite{Furui:2005bu,Furui:2003jr,Cucchieri:1997dx}.
We investigate the infrared behavior of $\phi(p)$ from our data in Section 
\ref{res} below. There we also comment on the approximation
obtained by neglecting terms with $p\neq q$ in Eq.\ (\ref{eq:phimodul}).

As already stressed in the Introduction, the dependence of $\phi(p)$ on
${\widetilde A}(q)$ [see Eq.\ (\ref{eq:phimom})] could explain
the $1/L^2$ behavior observed for 
$\langle \, \left| \, \phi(p) \, \right| \, \rangle$.
On the other hand, as mentioned before, 
this result does not imply that $\phi(p)$ is zero
at infinite volume.

Finally, from Eq.\ (\ref{eq:GeofGo}) we obtain that the off-diagonal
symmetric components of $\langle \, G_e^{bc}(p) \, \rangle$ should also be
non-zero if the off-diagonal anti-symmetric components are non-vanishing.
However, in this case --- due to the factor $k_{\mu} \, {\widetilde A}^{h}_{\mu}(p - k) / p^2$
--- one should expect a more singular dependence
on the inverse momentum $1/p$ and
stronger finite-volume effects than those observed for the anti-symmetric
components $\langle \, G_o^{bc}(p) \, \rangle$.
 

\subsection{The ghost-gluon vertex}
\label{sec:ggg}

Following the notation in Refs.\ \cite{Cucchieri:2004sq,Mihara:2004bx},
the ghost-gluon vertex function is given (on the lattice) by
the relation
\beq
\Gamma^{bcd}_{\!\mu}(\hat{k},\hat{p}) \; = \;
     \frac{V\,\langle\,{\widetilde A}_{\mu}^b(\hat{k})\,
                                     G^{cd}(\hat{p})  \,
                                   \rangle}{
\langle\,D(\hat{k}) \,\rangle\,\langle\,G(\hat{s})\,\rangle\,\langle\, G(\hat{p})\,\rangle} \,\mbox{,}
\label{eq:gammafin}
\eeq
where $\widetilde{s} = \widetilde{k}+\widetilde{p}$, $\langle\,D(\hat{k}) \,\rangle$
is the lattice gluon propagator and ${\widetilde A}_{\mu}^b(\hat{k})$ is defined in
Eq.\ (\ref{eq:Ak0}).
At tree-level (on the lattice) one has \cite{Kawai:1980ja}
\beq
\Gamma_{\mu}^{bcd}(\hat{k},\hat{p})
          \; =\; i \, g_0 \, f^{bcd} \hat{p}_{\mu}
\cos\left(\frac{\pi\,\widetilde{s}_{\mu}\, a}{L_{\mu}}\right)
\,\mbox{.}
\label{eq:lamb0}
\eeq
Thus, we can write
\beq
\Gamma_{\mu}^{bcd}(\hat{k},\hat{p})
          \; =\; i \, g_0 \, f^{bcd} \hat{p}_{\mu}
\cos\left(\frac{\pi\,\widetilde{s}_{\mu}\, a}{L_{\mu}}\right) \, \Gamma(\hat{k},\hat{p})
\eeq
and at the asymmetric point, with zero momentum for the gluon ($\hat{k} = 0$),
we find \cite{Cucchieri:2004sq}
\beqa
  \!\!\!\!\!\!\! \Xi(\hat{p}) &=&
   \frac{-i}{g_0 \, N_c  \,(N_c^2 \, - \, 1)} \frac{1}{\hat{p}^2} \,\sum_{\mu}\; \frac{\hat{p}_{\mu}}{
              \cos\left(\frac{\pi \,\widetilde{p}_{\mu}\, a}{L_{\mu}}\right)} \nonumber\\[2mm]
                    & & \qquad \;\;\;\;\;\; \times \;  \frac{V\,f^{bcd}\,\langle\,{\widetilde A}_{\mu}^b(0)\,
                                     G^{cd}(\hat{p})  \, \rangle}{
\langle\,D(0) \,\rangle\,\langle\,G(\hat{p})\,\rangle^2} \,\mbox{,}
\label{eq:xilattice}
\eeqa
where we set $\Xi(\hat{p}) = \Gamma(0,\hat{p})$.
Note that the combination $f^{bcd} G^{cd}(\hat{p}) / N_c$ is the same
that appears in the definition of
$\phi^b(\hat{p})$. This gives
\beqa
  \!\!\!\!\!\!\! \Xi(\hat{p}) &=&
   \frac{-i}{g_0 \,(N_c^2 \, - \, 1)} \frac{1}{\hat{p}^2} \,\sum_{\mu}\; \frac{\hat{p}_{\mu}}{
              \cos\left(\frac{\pi \,\widetilde{p}_{\mu}\, a}{L_{\mu}}\right)} \nonumber\\[2mm]
                    & & \qquad \;\;\;\;\;\; \times \;  \frac{V\,\langle\,{\widetilde A}_{\mu}^b(0)\,
                                     \phi^{b}(\hat{p})  \, \rangle}{
\langle\,D(0) \,\rangle\,\langle\,G(\hat{p})\,\rangle^2} \,\mbox{.}
\label{eq:xilattice2}
\eeqa
On the other hand, since in the above equation the quantity $\phi^{b}(\hat{p})$
is multiplied by
${\widetilde A}_{\mu}^b(0)$, one gets that the scalar function $\Xi(\hat{p})$
does not change sign under the transformation (\ref{eq:AtomA}).
This explains why there is no sign problem \cite{Cucchieri:2004sq,Mihara:2004bx}
when evaluating the ghost-gluon-vertex renormalization function
$\Z1^{-1}(\hat{p}) \; = \; \Xi(\hat{p})$.

Let us also note that the factor $\,\cos\left(\pi \,\widetilde{p}_{\mu}\, a/L_{\mu}\right)$, which
appears in the tree-level result (\ref{eq:lamb0}) and in the denominator of Eq.\ (\ref{eq:xilattice2}),
is a discretization effect that goes to 1 in the formal continuum limit $a \to 0$.
Due to the relation between $\Xi(\hat{p})$ and $\phi^{b}(\hat{p})$ [see Eq.\ (\ref{eq:xilattice2})],
it is likely that similar discretization effects come up also in the present work.
This will be verified in the next section.


\section{Results}
\label{res}

\begin{figure}[t]
\begin{center}
\protect\hskip -1.2cm
\protect\vskip 0.4cm
\includegraphics[height=0.80\hsize]{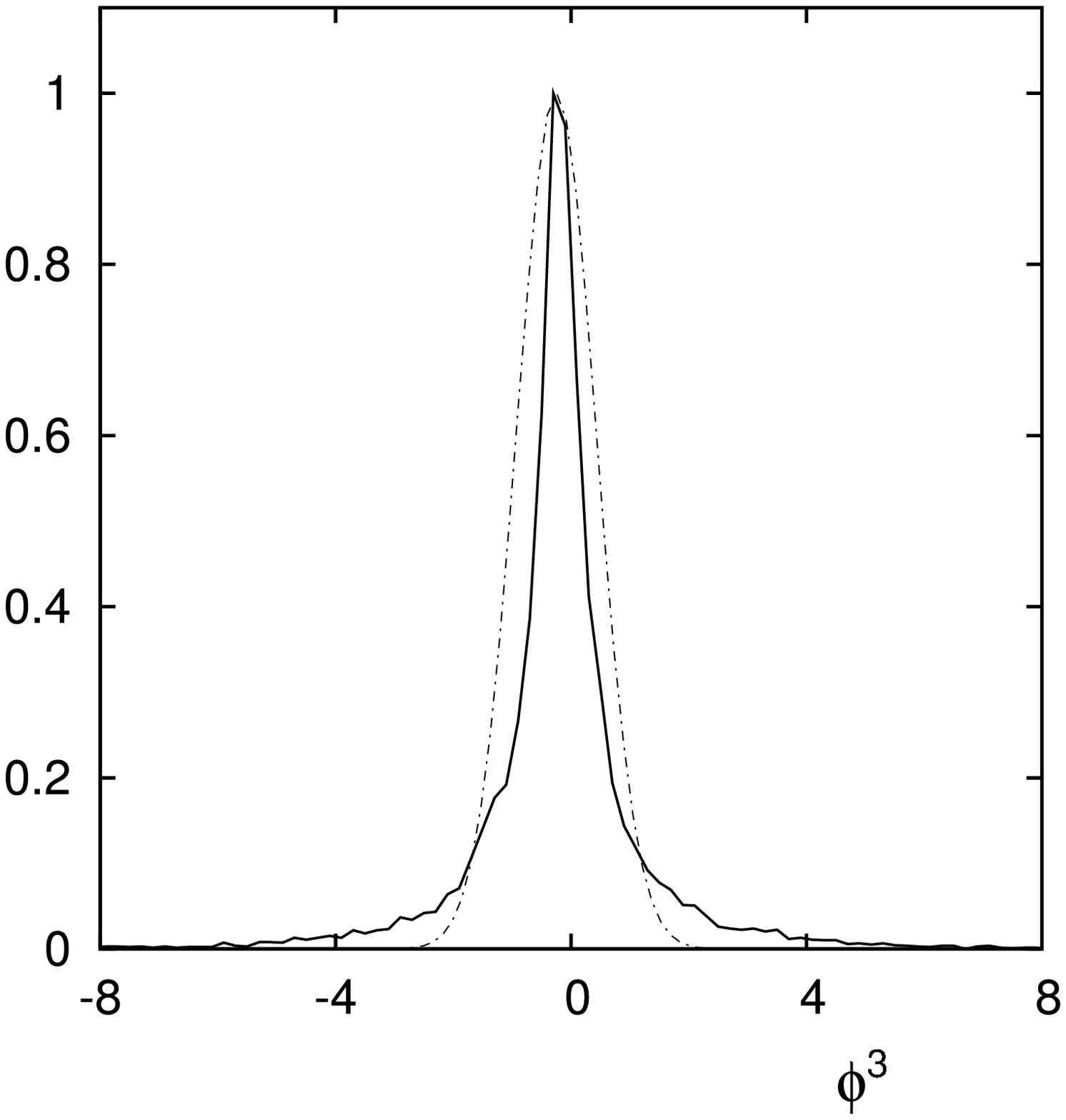}
\caption{Histogram (solid curve)
of $\phi^3(\hat{p})$
for the lattice volume $8^4$ with $\beta = 2.4$
and $\widetilde{p} = (0, 0, 0, 1)$, using 10,000 configurations.
For comparison we also plot a Gaussian distribution
(dot-dashed line) with
the same mean value and the same standard deviation.
\label{fig:histphi}
}
\end{center}
\end{figure}

\begin{figure}[b]
\begin{center}
\protect\hskip -1.2cm
\protect\vskip 0.4cm
\includegraphics[height=0.80\hsize]{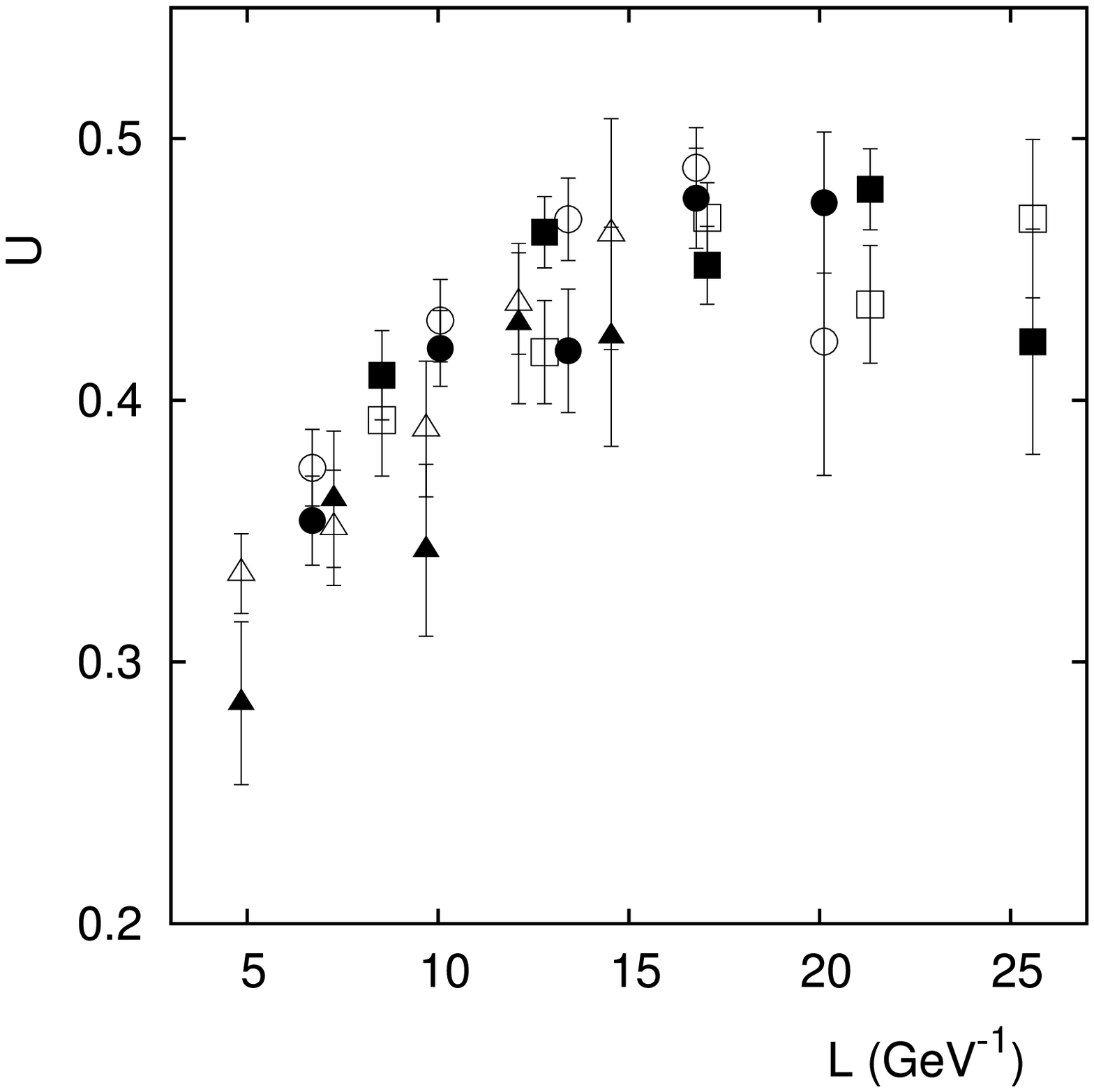} 
\caption{Results for the Binder cumulant $U$ [defined in Eq.\ (\ref{eq:binder})]
for the quantity $\phi^b(\hat{p})$ as a function of the lattice
side $L$ (in GeV$^{-1}$) for all lattice volumes $V$, couplings $\beta$ and
momenta with $\widetilde{p} = N/4$.
We show the data corresponding to asymmetric momenta [for $\beta = 2.2$
($\Box$), $\beta = 2.3$ ($\bigcirc$) and $\beta = 2.4$ ($\bigtriangleup$)]
and to symmetric momenta (with the corresponding filled symbols
for each $\beta$).
Errors have been estimated using the bootstrap method with 10,000 samples.
\label{fig:binder}
}
\end{center}
\end{figure}

We start our analysis of the data by considering the two Fourier-transformed
quantities
\beqa
G_e^{bd}(\hat{p}) \!\!&=&\!\! \frac{1}{V} \, \sum_{x\, y} \,
             G^{bd}(x\mbox{,}y) \,
            \cos{\left[\,\widetilde{p}\, (x - y)\,\right]} \\[2mm]
G_o^{bd}(\hat{p}) \!\!&=&\!\! \frac{1}{V} \, \sum_{x\, y} \,
             G^{bd}(x\mbox{,}y) \,
            \sin{\left[\,\widetilde{p}\, (x - y)\,\right]}
\, \mbox{.}
\eeqa
As explained in Section \ref{cond}, we expect $G_e^{bd}(\hat{p})$ [respectively
$G_o^{bd}(\hat{p})$] to be symmetric (respectively anti-symmetric) in the color
indices $b$ and $d$. We have verified this by evaluating, for {\em each}
configuration and for {\em each} momentum $\hat{p}$, the quantities
\beqa
& &\left| \, \frac{ G_e^{bd}(\hat{p}) \,-\, G_e^{db}(\hat{p}) }{
        \left[ G_e^{bd}(\hat{p}) \,+\, G_e^{db}(\hat{p}) \right] / 2}\,\right|
\\[2mm] 
& &\left| \, \frac{ G_o^{bd}(\hat{p}) \,+\, G_o^{db}(\hat{p}) }{
        \left[ G_o^{bd}(\hat{p}) \,-\, G_o^{db}(\hat{p}) \right] / 2} \,\right|
\,\mbox{,}
\eeqa
which should both be zero. We find that these two quantities are
usually very small, being respectively of the order of $10^{-3}$ and $10^{-6}$.
At the same time, we checked the hypothesis of strong diagonal-dominance
for the symmetric components $G_e^{bd}(\hat{p})$. We obtain that the relation
\beq
 | \, G_e^{bb}(\hat{p}) \, | \; > \; | \, \sum_{d \neq b} \, G_e^{bd}(\hat{p}) \, |
\eeq
is satisfied for {\em almost all} momenta $\hat{p}$ and configurations
(except for 14 cases out of 280800). This is expected since, due
to the global color symmetry and for a zero ghost condensate $v$, one should
have $\langle \, G^{ed}_e(\hat{p}) \, \rangle \propto \delta^{ed}$, after averaging over the
gauge-fixed Monte Carlo configurations. Thus, it seems reasonable to assume that
$G^{ed}_e(\hat{p})$ is strongly diagonally-dominant already for each configuration. This should
also happen if the ghost condensate $v$ is nonzero and small.
We also considered the propagator $G_e^{bd}(\hat{k},\hat{p})$ with $\hat{k} \neq \hat{p}$
for the lattice volume $8^4$ and $\beta = 2.4$ with $\widetilde{k}$ given by
$(0, 0, 0, 1)$ in the asymmetric case and by $(1, 1, 1, 1)$ in the
symmetric one. In this case we find that the
relation   
\beq
 | \, G_e^{bb}(\hat{k}, \hat{p}) \, | \; > \; | \, \sum_{d \neq b} \, 
          G_e^{bd}(\hat{k}, \hat{p}) \, |
\eeq
is satisfied in about $75 \%$ of the cases.

We then evaluated the average of [see Eq.\ (\ref{eq:iphi})]
\beq
i \, \phi^b(\hat{p}) \;=\; \frac{i \, \epsilon^{bcd}}{2} \, G^{cd}(\hat{p}) \; = \; 
  \frac{\epsilon^{bcd}}{2} \, G_o^{cd}(\hat{p})
\eeq
for $b = 1, 2, 3$. We find that this quantity is consistent with zero, within
one standard deviation~\footnote{All the errors have been evaluated using the bootstrap method
with 10,000 samples. We checked that the errors calculated in this way are in
agreement with the estimates obtained when considering one standard deviation.}.
Indeed, the statistical fluctuations are always very large, being larger
than the central value for more than $60 \%$ of the data points and larger
than $1/3$ of the central value for all our data.
On the contrary, when considering the absolute value of $ \, \phi^b(\hat{p}) \, $
we find a relatively small error, of less than $10 \%$ of the central
value in $99.7 \%$ of the cases. As discussed in the Introduction,
this is a clear indication that the data for $ \, \phi^b(\hat{p}) \, $
do not correspond to a Gaussian distribution. We have verified this
by comparing a histogram of the data with a Gaussian curve.
Then, as for the $3d$ $O(4)$ nonlinear $\sigma$-model in the broken phase
(see the Introduction), one finds (see Fig.\ \ref{fig:histphi}) that the 
shape of the statistical distribution of the data is clearly not Gaussian. 
(However, notice that the shape of this non-Gaussian distribution is not 
the same as for the spin-model case.)
We also find that working at constant physics, i.e.\ with approximately fixed 
lattice side $L$ and momentum $p$ (in physical units),
the deviation from a Gaussian distribution increases
when the lattice coupling $\beta$ increases.
Finally, we have looked at the Binder cumulant $\,U\,$
[see Eq.\ (\ref{eq:binder})] for the
order parameter $\phi^b(\hat{p})$.
We find that $\,U\,$ is clearly different from zero
for all lattice sizes $N$, couplings $\beta$ and momenta $\hat{p}$, 
and seems to converge to a value $U \approx 0.45$ as the lattice side $L$ increases
(see Fig.\ \ref{fig:binder}).
This limiting value is essentially independent of
the momentum $p$~\footnote{We do not include in this analysis the momenta (asymmetric
and symmetric) with $\widetilde{p} = 1$ and with $\widetilde{p} = N/2$,
since they are often characterized by
a Binder cumulant $U$ with a very large statistical error.}.
The data show good scaling for the three $\beta$ values
considered.
As argued in the Introduction, the large fluctuations for $\phi^b(\hat{p})$ and the
deviation from a Gaussian distribution could signal the spontaneous breaking
of the global $SL(2,R)$ symmetry, implying a nonzero value for the
ghost condensate $v$.

\begin{figure}[t]
\begin{center}
\protect\hskip -1.2cm
\protect\vskip 0.4cm
\includegraphics[height=0.80\hsize]{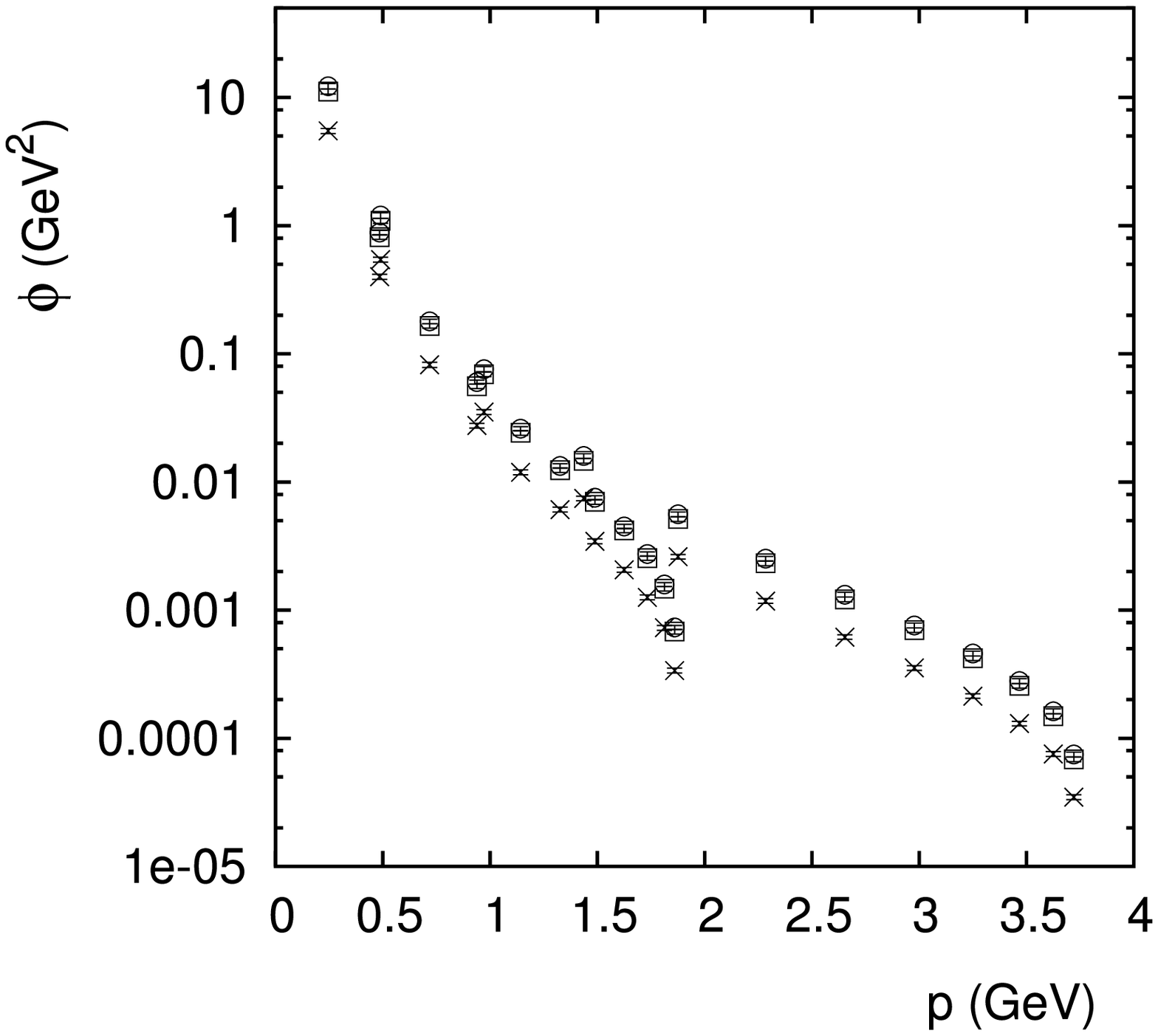} 
\caption{Results for $\phi_a$ ($\times$), $\phi_m$ ($\Box$)\ and $\phi_r$ ($\circ$)
[defined in Eqs.\
(\ref{eq:phia})--(\ref{eq:phir})] as a function of the momentum $p$
for lattice volume $V= 24^4$ and
$\beta = 2.2$. We show the data corresponding to asymmetric and
to symmetric momenta. Note the logarithmic scale on the $y$ axis.
Errors have been estimated using the bootstrap method with 10,000 samples.
\label{fig:phiabs}
}
\end{center}
\end{figure}

In order to plot the data for $ | \phi^b(\hat{p}) | $, we average over the color index $b$ considering the
three quantities
\beqa
\phi_{a}(\hat{p}) &=& \frac{1}{3} \,\sum_{b} \, \langle \, | \, \phi^b(\hat{p}) \, | \,\rangle \label{eq:phia} \\[2mm]
\phi_{m}(\hat{p}) &=& \langle \,\sqrt{ \, \frac{1}{3} \,\sum_{b}\, | \, \phi^b(\hat{p}) \, |^2 \,} \,\rangle \label{eq:phim} \\[2mm] 
\phi_{r}(\hat{p}) &=& \sqrt{ \,\langle \,\frac{1}{3} \,\sum_{b}\, | \, \phi^b(\hat{p}) \, |^2 \,\rangle} \label{eq:phir}
\,\mbox{,}
\eeqa
i.e.\ we take (respectively) the average of the absolute value of the components of $\phi$, the average
of the absolute value of $\phi$ and the root mean square of $\phi$. The results are shown
in Fig.\ \ref{fig:phiabs}. In this figure the breaking of rotational symmetry is
evident: the asymmetric and symmetric momenta clearly form two different curves. In
particular, the three quantities decrease significantly at $p \approx 1.85$ GeV
in the asymmetric case and at $p \approx 3.7$ GeV in the symmetric one.

\begin{figure}[t]
\begin{center}
\protect\hskip -1.2cm
\protect\vskip 0.4cm
\includegraphics[height=0.80\hsize]{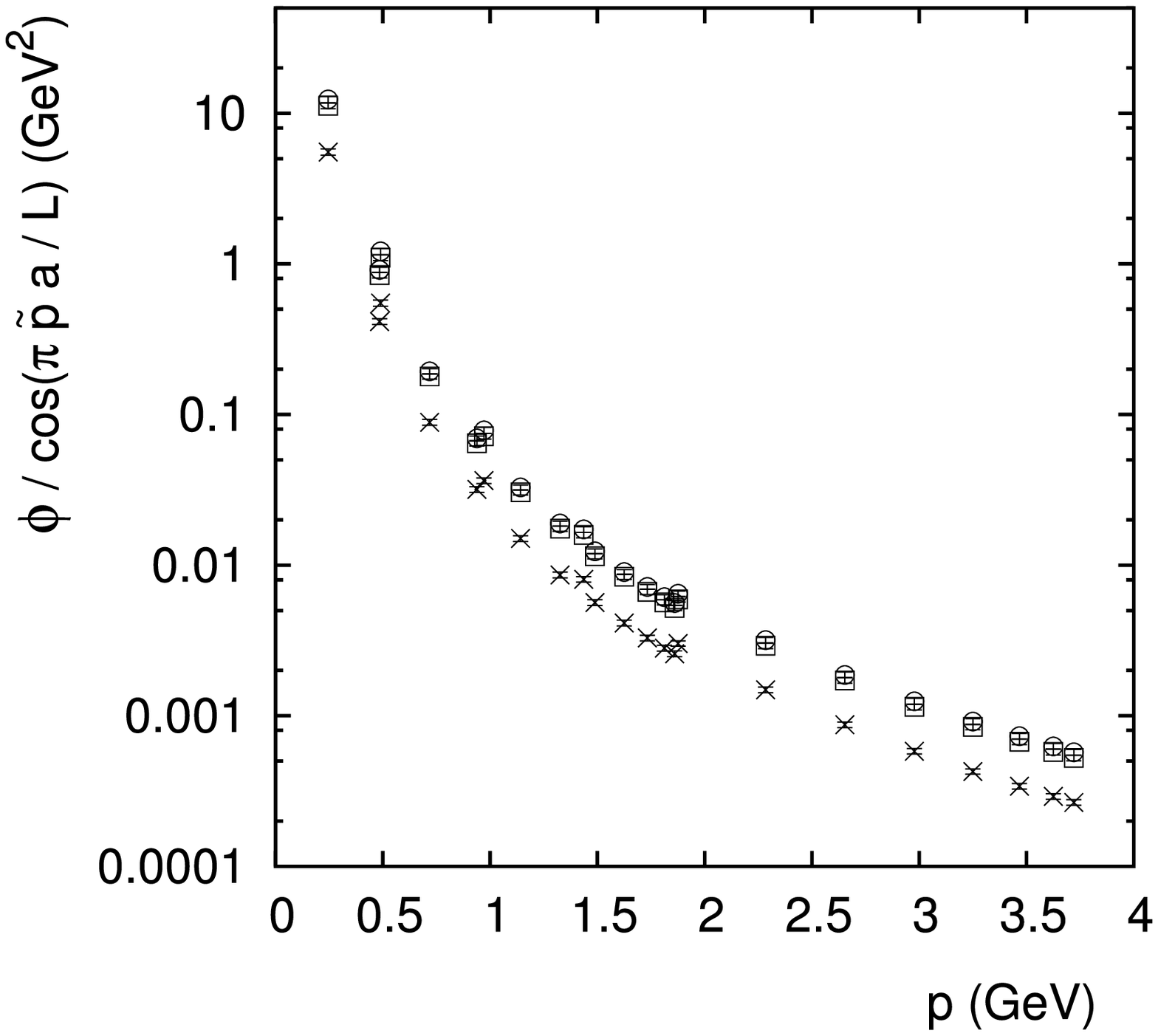}
\caption{Results for $\phi_a$ ($\times$), $\phi_m$ ($\Box$)\ and $\phi_r$ ($\circ$)
[defined in Eqs.\
(\ref{eq:phia})--(\ref{eq:phir})], divided by
$\cos\left(\pi \,\widetilde{p}\, a/L\right)$,
as a function of the momentum $p$
for lattice volume $V= 24^4$ and $\beta = 2.2$.
We show the data corresponding to asymmetric and to symmetric momenta.
Note the logarithmic scale on the $y$ axis.
Errors have been estimated using the bootstrap method with 10,000 samples.
\label{fig:phiabsrisc}
}
\end{center}
\end{figure}

The interpretation of the data is simpler (see Fig.\ \ref{fig:phiabsrisc})
if one rescales $\phi^b(\hat{p})$ with the factor
$\cos\left(\pi \,\widetilde{p}\, a/L\right)$ that appears in the
lattice formula for the ghost-gluon-vertex renormalization function [see Eq.\ (\ref{eq:xilattice2})
in Section \ref{sec:ggg}]. Moreover, as can be seen in Fig.\ \ref{fig:phiabsrisc},
the effects due to the breaking of rotational symmetry are relatively small.
Finally, we can multiply the data by $L^2$ to obtain the single curve shown in
Fig.\ \ref{fig:phiabsfin}.
Again, we find good scaling for the three $\beta$ values
considered. Let us recall that, for the gluon and ghost
propagators in the $SU(2)$ case \cite{Bloch:2003sk,Bloch:2002we},
scaling has been observed in the range $\beta \in [2.15, 2.8]$.
Note that, from our analysis in Section \ref{num}, we would
expect (in the infrared limit)
a behavior of the type $\,G(p)/ p\,$ for the three quantities defined above,
if the contribution proportional to ${\widetilde A}(0)$ is dominant.

\begin{figure}[t]
\begin{center}
\protect\hskip -1.2cm
\protect\vskip 0.8cm
\includegraphics[height=0.80\hsize]{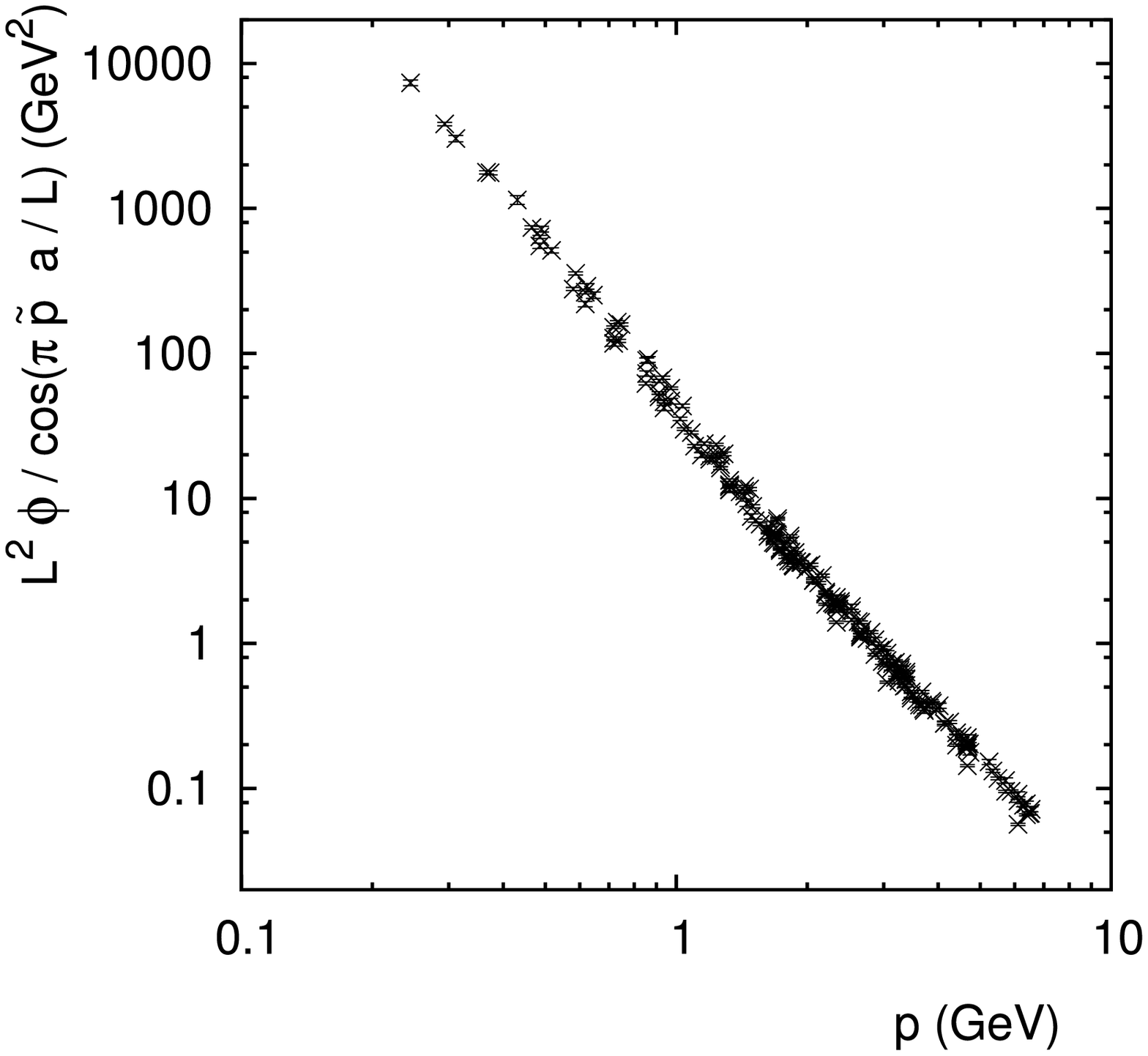}
\caption{Results for $\phi_m$ [defined in Eq.\ (\ref{eq:phim})], rescaled by
$L^2 / \cos\left(\pi \,\widetilde{p}\, a/L\right)$,
as a function of the momentum $p$ for all lattice volumes and $\beta$ values considered.
We show the data corresponding to asymmetric and to symmetric momenta.
Note the logarithmic scale on the $x$ and $y$ axes.
Errors have been estimated using the bootstrap method with 10,000 samples.
\label{fig:phiabsfin}
}
\end{center}
\end{figure}

Let us recall that in the $SU(2)$ case the inverse Faddeev-Popov matrix
--- in the condensed Overhauser vacuum $\phi^b \propto \delta^{b3}$ ---
should be given (in momentum space) by \cite{Dudal:2003dp}
\beq
G^{cd}(p) \;=\; \frac{p^2 \, \delta^{cd}\, + \, v\,\epsilon^{cd}}
{p^4 + v^2} \,\mbox{,}
\label{fpm0}
\eeq
where $\epsilon^{cd}$ is the anti-symmetric tensor with $c$, $d = 1$, $2$ 
and $v$ is the ghost condensate, given (at one loop and for $N_c = 2$) by \cite{Dudal:2003dp}
\beq
v \;=\; \left| \frac{\phi^3}{\rho_0} \right| 
  \;=\; \left| \frac{\chi_{vac} \,\left( 2\,N_f\,-\,13 \right)}{3} \right|
\eeq
with $\chi_{vac} = 0.539 \, \Lambda_{\overline{MS}}^2\,$.
Since in our case $N_f = 0$, we find
\beq
v \;\approx \; \frac{13}{3}\,0.539 \, \Lambda_{\overline{MS}}^2 \;\approx\;
                             2.34\, \Lambda_{\overline{MS}}^2
\,\mbox{.}
\eeq
Then, with $\Lambda_{\overline{MS}} \approx 250$ MeV, we obtain
$v \approx 0.146$ GeV$^2$. A fit of the type
\beq
\frac{L^2}{\cos\left(\frac{\pi \,\widetilde{p}\, a}{L}\right)} 
\, |\,\phi^{b} \,| \;=\; \frac{r}{p^z} \,\mbox{,}
\label{eq:fitfunc}
\eeq
works quite well for our data, giving for the parameters $r$ and $z$ the results
reported in Table \ref{tab:fit}. Note that the exponent $z$ increases
systematically when one considers data at small momenta only, suggesting that
$z$ would approach $4$ if one could really probe the infrared limit.
We also tried a fit using the form $\,r / (p^z \,+\, v^2)$. The results
obtained in this case are very similar to those reported in Table \ref{tab:fit}, with
$\,v\,$ always consistent with zero within error bars.
Furthermore, for the ghost propagator $G(p)$ we find that a fit to
$\,r/p^z$ yields $z$ in the interval $2.33 - 2.46$, depending
on the set of data considered.
Thus, the quantity $\phi(p)$ has a stronger infrared behavior than $G(p)/p$, which
is obtained by neglecting terms with $p\neq q$ in Eq.\ (\ref{eq:phimodul}).

\begin{table}[t]
\begin{center}
\begin{tabular}{|c|c|c|c|c|}
\hline
Data considered                            &  d.o.f.   & $r$     & $ z $   & $ \chi^2/d.o.f.$ \\ \hline
All data                                   &  208      & 37.3(7) & 3.52(2) & 6.54 \\ \hline
$V=24^4$ and $\beta = 2.2$                 &   20      & 36(1)   & 3.65(5) & 3.09 \\ \hline
All data                                   &  96       & 39.5(6) & 3.79(3) & 5.33 \\
  and $p < 2$ GeV                          &           &         &         &      \\ \hline
$V=24^4$, $\beta = 2.2$                    &  13       & 36(1)   & 3.80(5) & 2.50 \\
  and $p < 2$ GeV                          &           &         &         &      \\ \hline
All data, $p < 2$ GeV,                     &  24       & 49(1)   & 3.82(5) & 3.44 \\
  only symm.\ momenta                      &           &         &         &      \\ \hline
All data, $p < 2$ GeV,                     &  70       & 37.7(4) & 3.80(2) & 3.35 \\
  only asymm.\ momenta                     &           &         &         &      \\ \hline
$V=24^4$, $\beta = 2.2$,                   &  9        & 34.3(4) & 3.80(2) & 0.82 \\
  only asymm.\ momenta                     &           &         &         &      \\ \hline
\end{tabular}
\caption{The parameters $r$ and $z$ obtained by fitting $\phi_{m}$ to
the formula (\ref{eq:fitfunc}). We considered different sets of data.
The fits have been done using {\tt gnuplot}.
\label{tab:fit}}
\end{center}
\end{table}

By comparing the above fitting formula to the
theoretical prediction (\ref{fpm0}), i.e.\ by considering
(in the anti-symmetric case) the Ansatz \footnote{Note that here we
neglect the discretization effect related to the factor 
$\cos\left(\pi \,\widetilde{p}\, a/L\right)$.}
\beq
G^{cd}(p) \;=\; \frac{r \, L^{-2} \,+\,v}{p^4 + v^2}\,\epsilon^{cd} \,\mbox{,}
\label{fpm0mod}
\eeq
we can state that
\beq
v \,\ll\, r\,L^{-2}_{max}\,, \quad v \,\ll\, p^2_{min} \,\mbox{.}
\label{eq:vineq}
\eeq
In our simulations we have $1/L_{max} \approx 0.039$ GeV
and $p_{min} \approx 0.245$ GeV.
Thus, using $r \approx 38$ (see Table \ref{tab:fit}), 
we obtain our final estimate 
\beq
v \ll 0.058 \;  \mbox{GeV}^2
\,\mbox{,}
\eeq
Note that, since $38 \approx (2 \pi)^2$ and $p_{min} \approx 2 \pi / L_{max}$,
the two inequalities in (\ref{eq:vineq}) give practically
the same estimate for the ghost condensate $v$.
One can also try a fit of the data using the function $r / (p^z \,+\, v^2)$ with
a fixed small value for $\,v\,$. Then, for $v \leq 0.025 \, \mbox{GeV}^2$ the
values of $ \chi^2/d.o.f.$ reported in Table \ref{tab:fit} do not change
significantly.

\begin{figure}[t]
\begin{center}
\protect\hskip -1.2cm
\protect\vskip 0.4cm
\includegraphics[height=0.80\hsize]{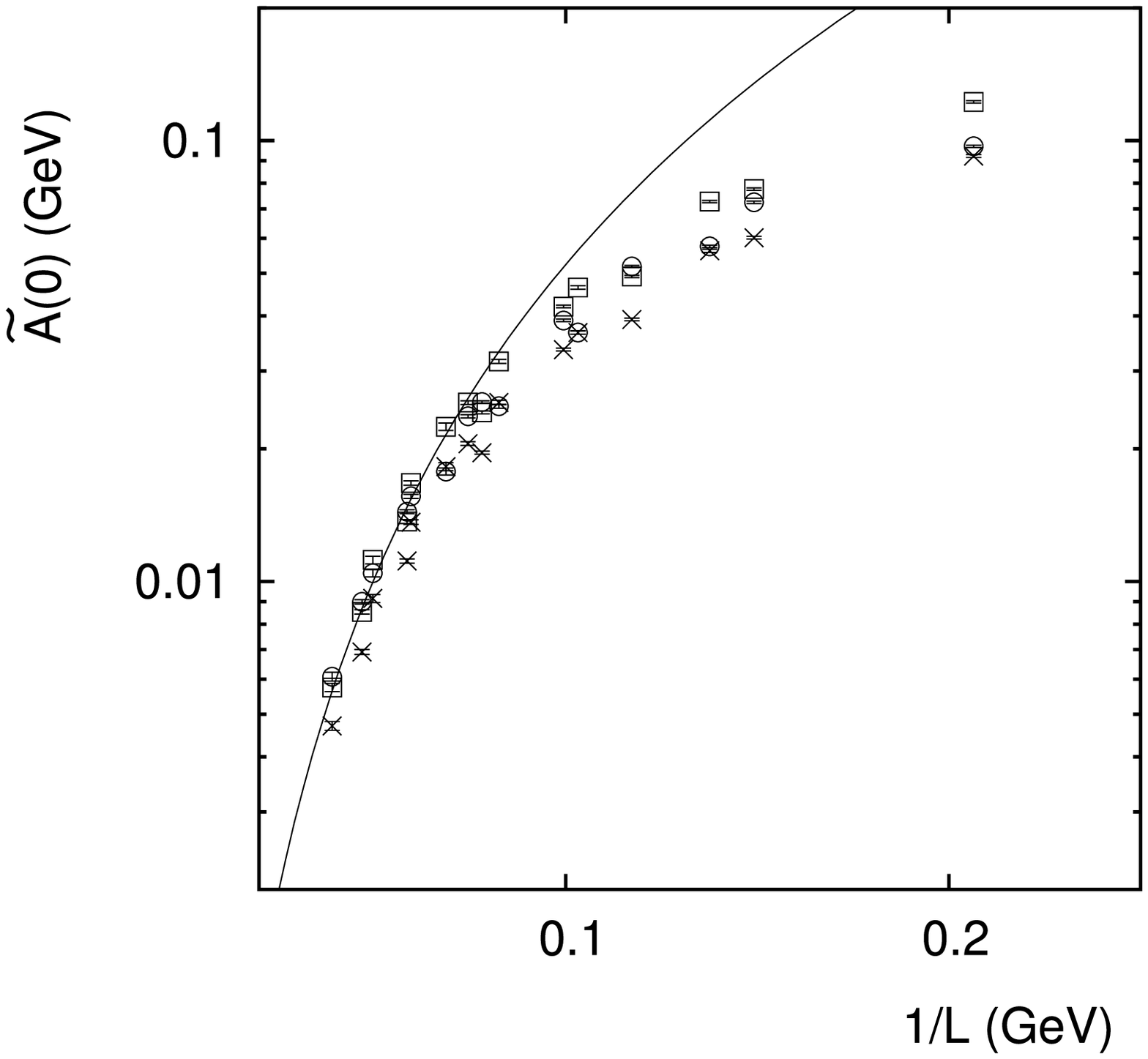}
\caption{Results for ${\widetilde A}_{a}(0)$ ($\times$), ${\widetilde A}_{m}$ ($\Box$)\ 
and ${\widetilde A}_{r}(0)$ ($\circ$)
[defined in Eqs.\ (\ref{eq:Aa})--(\ref{eq:Ar})] as a function of the inverse
lattice side $1/L$
for all lattice volumes and $\beta$ values considered.
Errors have been estimated using the bootstrap method with 10,000 samples.
We also report the fit of ${\widetilde A}_{m}$
to the function $b / L^s$ with $b = 10(7)$ and $s = 2.4(2)$.
When considering the data for ${\widetilde A}_{a}(0)$, one finds for the exponent $s$ the
same value as above, while for ${\widetilde A}_{r}(0)$ one gets
$s = 2.0(1)$. In all cases, the fits have been done using
{\tt gnuplot} and we consider only the data corresponding to $1/L < 0.07$ GeV.
\label{fig:Aabs}
}
\end{center}
\end{figure}

\begin{figure}[t]
\begin{center}
\protect\hskip -1.2cm
\protect\vskip 0.4cm
\includegraphics[height=0.80\hsize]{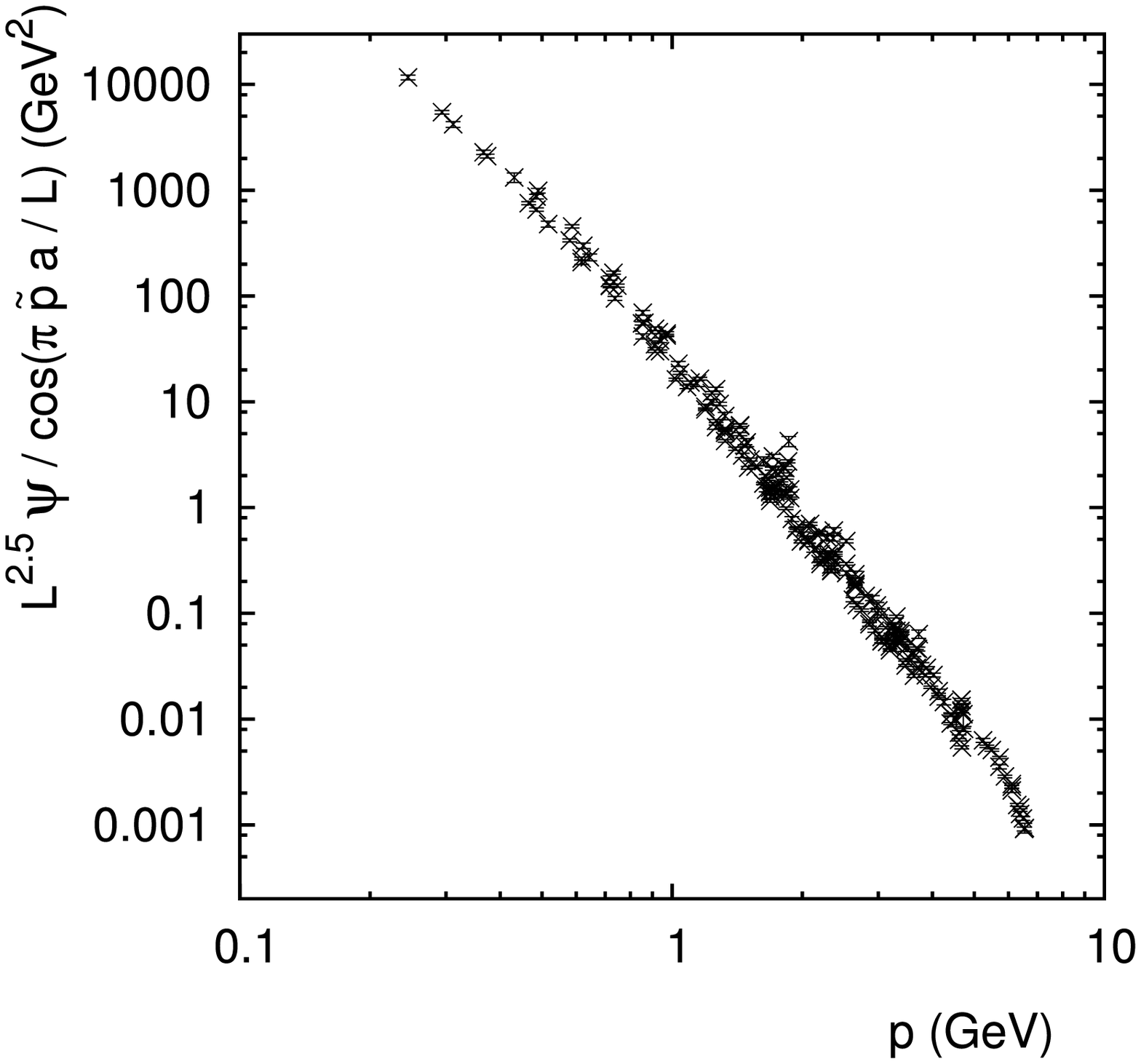}
\caption{Results for $\psi_m$ [defined in Eq.\ (\ref{eq:psim})], rescaled by
$L^{2.5} / \cos\left(\pi \,\widetilde{p}\, a/L\right)$,
as a function of the momentum $p$ for all lattice volumes and $\beta$ values considered.
We show the data corresponding to asymmetric and to symmetric momenta.
Note the logarithmic scale on the $x$ and $y$ axes.
Errors have been estimated using the bootstrap method with 10,000 samples.
\label{fig:psiabsfin}
}
\end{center}
\end{figure}

Now, we consider the dependence of $| {\widetilde A}_{\mu}^b(q) |$ on the
lattice side $L$ in physical units. In analogy with Eqs.\ (\ref{eq:phia})--(\ref{eq:phir}),
we define the three quantities
\beqa
{\widetilde A}_{a}(q) &=& \frac{1}{12} \,\sum_{b} \, \sum_{\mu} \langle \, | \, {\widetilde A}_{\mu}^b(q) \, | 
              \,\rangle \label{eq:Aa} \\[2mm]
{\widetilde A}_{m}(q) &=& 
  \langle \,\sqrt{ \, \frac{1}{12} \,\sum_{b}\, | \, {\widetilde A}_{\mu}^b(q) \, |^2 \,} \,\rangle \label{eq:Am} \\[2mm]
{\widetilde A}_{r}(q) &=& \sqrt{ \,\langle \,\frac{1}{12} \,\sum_{b}\, | \, {\widetilde A}_{\mu}^b(q) \, |^2 
              \,\rangle} \label{eq:Ar}
\,\mbox{.}
\eeqa
As shown in Fig.\ \ref{fig:Aabs}, we find that the (absolute value of the)
Fourier-transformed gluon field at zero momentum ${\widetilde A}(0)$ goes
to zero as $L^{-s}$ with $s \approx 2$ for the three quantities defined above.
Similar results can be obtained when $q \neq 0\,$~\footnote{Of course, for a given
physical value of the momentum $q \neq 0$, we have that only a few values of $L$
can be considered for this fit.}.
This is a confirmation that the finite-size effects
observed for the quantity $\phi(p)$ are indeed induced by ${\widetilde A}(q)$.
We have also checked that the sign of $\phi^3(p)$ can be changed using the global
gauge transformation reported in Eq.\ (\ref{eq:gsign}), i.e.\ by changing the sign of
${\widetilde A}_{\mu}^3(q)$.

Finally, we study the off-diagonal symmetric components $G_e^{bc}(x\mbox{,}y)$
by considering the quantity $\,\psi^b(\hat{p}) \,=\,\left[\,
G_e^{cd}(\hat{p}) \,+\,G_e^{dc}(\hat{p})\,\right] / 2\,$,
with $b \neq c, d$ and $b = 1$, $2$, $3$.
Let us stress that, even though these components are symmetric under the
transformation (\ref{eq:AtomA}), they are not necessarily positive
[see Eq.\ (\ref{eq:Gsymmetric})].
Thus, also in this case we should consider their absolute value.
In analogy with the analysis reported above for $\phi^b(\hat{p})$, we can evaluate
\beqa
\psi_{a}(\hat{p}) &=& \frac{1}{3} \,\sum_{b} \, \langle \, | \, \psi^b(\hat{p}) \, | \,\rangle \label{eq:psia} \\[2mm]   
\psi_{m}(\hat{p}) &=& \langle \,\sqrt{ \, \frac{1}{3} \,\sum_{b}\, | \, \psi^b(\hat{p}) \, |^2 \,} \,\rangle \label{eq:psim} \\[2mm]
\psi_{r}(\hat{p}) &=& \sqrt{ \,\langle \,\frac{1}{3} \,\sum_{b}\, | \, \psi^b(\hat{p}) \, |^2 \,\rangle} \label{eq:psir}
\,\mbox{.}
\eeqa
We obtain that these quantities produce a single
curve when rescaled by $L^{2.5} /
\cos\left(\pi \,\widetilde{p}\, a/L\right)$ 
(see Fig.\ \ref{fig:psiabsfin}). Note however that the rescaling does not work
as well as for the $\phi$'s.
Also, the rescaled data can be reasonably well fitted using the fitting
function $ \, s / p^z\,$
with $z \approx 5$. This is in agreement with the observation made
before Section \ref{sec:ggg} and in partial agreement with the results
reported in Ref.\ \cite{Furui:2003jr}.

We have done the analysis reported above also for the gauge-fixed
configurations obtained using the smearing method \cite{Hetrick:1997yy}.
As for the ghost-gluon
vertex \cite{Cucchieri:2004sq,Mihara:2004bx},
we find that the contribution of Gribov-copy effects (if present)
is always zero within error bars. In particular, the exponent $z$ obtained when fitting
the functions $\phi(p)$ is close to 4 at small momenta also
when using the smearing method.


\section{Conclusions}
\label{con}

We study numerically the off-diagonal components of the momentum-space ghost propagator
$G^{cd}(p)$ for $SU(2)$ lattice gauge theory in minimal Landau
gauge. We see clear signs of spontaneous breaking of a global symmetry,
using the quantity $\phi^b(p) = \epsilon^{bcd}
G^{cd}(p) / 2$ as order parameter.
As in the case of continuous-spin models in the
ordered phase, we find indication of spontaneous symmetry breaking from
two (related) observations: 1) by comparing the statistical 
fluctuations for the quantities $\phi^b (p)$ and $|\phi^b (p)|$;
2) from the non-Gaussian shape of the 
statistical distribution of $\phi^b (p)$, which can be observed
by considering a histogram of the data or by evaluating the
so-called Binder cumulant.
Since, in Landau gauge, the vacuum expectation value of the
quantity $\phi^b (p)$ should be proportional (in the so-called
Overhauser channel) to the ghost
condensate $v$ \cite{Dudal:2003dp}, it seems reasonable
to conclude that the broken symmetry is the $SL(2,R)$ symmetry,
which is related to ghost condensation \cite{Dudal:2002ye,Alkofer:2000wg}.
Let us note that, from our data, the Binder cumulant $U$ seems
to be approximately null at small lattice volume and
to converge to a value $U \approx 0.45$ for physical lattice side $L \gtapprox 15$
Gev$^{-1} \approx 3$ fm, corresponding to a mass scale of less than $100$ MeV.

As stressed in the Introduction, we have shown that the dependence
of $\phi^b(p)$ on ${\widetilde A}^d_{\mu}(q)$ [see Eq.\ (\ref{eq:phimodul})] can explain
the $1/L^2$ behavior observed for $\langle \, \left| \, \phi(p) \, \right| \, \rangle$.
We have also shown that $\phi^b (p)$ has discretization effects similar to
those obtained for the ghost-gluon vertex \cite{Cucchieri:2004sq,Mihara:2004bx}.
Then, using the rescaled quantity
$|\, L^2 \, \phi^b (p) / \cos(\pi\tilde{p} a / L) \,|$,
we find a behavior $p^{-z}$ with
$z\approx 4$, in agreement with analytic predictions \cite{Dudal:2003dp}.
It is important to note that this result is not obtained just by considering
the approximation $\phi(p) \approx G(p) / p$ [see Eq.\ (\ref{eq:phiGp})],
since our data for the ghost propagator $G(p)$ have an infrared
behavior given by $1/p^{2.4}$. We therefore view the similar
momentum dependence obtained numerically and analytically for $\phi^b (p)$ 
as a further indication of ghost condensation.
On the other hand, from our fits we find
that the ghost condensate $v$ is consistent with zero within error bars, i.e.\
the quantity $|\,L^2 \phi^b (p) / \cos(\pi\tilde{p} a / L) \,|$ does not
approach a finite limit at small momenta, at least for $p \geq 0.245\,$ GeV.
Using the Ansatz (\ref{fpm0mod}) we
obtain for the ghost condensate the upper bound $v \ll 0.058$ GeV$^2$.
More precisely, our data rule out values of $v$ greater than
$0.058$ GeV$^2 \approx (240 $ MeV$)^2$ but would still be
consistent with a ghost condensate
$v \ltapprox 0.025$ GeV$^2 \approx (160 $ MeV$)^2$.
As said in the Introduction, in analytic studies
\cite{Dudal:2002xe,Dudal:2003dp,Capri:2005,Sawayanagi:2003dc} one finds that
the ghost condensate induces a tachyonic gluon mass proportional to
$\sqrt{v}$, which modifies the dynamic gluon mass related
to the gauge condensate $\langle A_{\mu}^b A_{\mu}^b \rangle$. 
Thus, one should expect a relatively small ghost condensate
$\,v\,$ in order to obtain a global (non-tachyonic) gluon mass.
Let us recall that a dynamic gluon mass of the order of a few
hundred MeV has been considered in several phenomenological
studies \cite{Parisi:1980jy,Halzen:1992vd,GayDucati:1993fn,Field:2001iu,GayDucati:2005pz}.
A similar mass scale was also obtained in
numerical studies of the gluon propagator in Landau gauge
\cite{Cucchieri:1999sz,Langfeld:2001cz,Cucchieri:2003di,Cucchieri:2004mf}.

We plan to extend our simulations to
larger lattice volumes, up to $48^4$, which
would allow us to detect a ghost condensate as small as
$v \approx 0.01\, $GeV$^2 =  (100 $ MeV$)^2$.  


\section*{ACKNOWLEDGMENTS}

We thank S.\ P.\ Sorella and R.\ F.\ Sobreiro
for many helpful discussions
and the IF-UERJ in Rio de Janeiro for the kind hospitality during 
our visits there.
We also thank S.\ Furui, H.\ Nakajima, M.\ Schaden and D.\ Zwanziger for 
useful comments and suggestions on the manuscript 
and D.\ Dudal for sending us Ref.\ \cite{Capri:2005} before submittal
to {\tt arXiv.org}. This research was supported by FAPESP 
(Projects No.\ 00/05047--5 and 03/00928--1).
Partial support from CNPq is also acknowledged (AC, TM).



\clearpage


\end{document}